\documentclass[lettersize,journal]{IEEEtran}
\usepackage{amsmath,amssymb,amsfonts}
\usepackage{algorithm}
\usepackage{algpseudocode}
\usepackage{array}
\usepackage{textcomp}
\usepackage{stfloats}
\usepackage{url}
\usepackage{verbatim}
\usepackage{graphicx}
\usepackage{cite}
\usepackage{subcaption}
\usepackage{multirow}
\usepackage{color}
\usepackage[colorlinks,linkcolor=blue]{hyperref}

\begin{document}

\title{MoNetV2: Enhanced Motion Network for Freehand 3D Ultrasound Reconstruction}

\author{Mingyuan Luo, Xin Yang, Zhongnuo Yan, Yan Cao, Yuanji Zhang, Xindi Hu, Jin Wang, Haoxuan Ding, Wei Han, Litao Sun, and Dong Ni
\thanks{This work was supported by the National Natural Science Foundation of China (Nos. 12326619, 62171290), Science and Technology Planning Project of Guangdong Province (No. 2023A0505020002), and Shenzhen Science and Technology Program (No. SGDX20201103095613036). (\textit{Corresponding author: Wei Han, Litao Sun, and Dong Ni.})}
\thanks{Mingyuan Luo, Xin Yang, Zhongnuo Yan, Yuanji Zhang, Haoxuan Ding, and Dong Ni are with the National-Regional Key Technology Engineering Laboratory for Medical Ultrasound, School of Biomedical Engineering, Shenzhen University Medical School, Shenzhen University, Shenzhen, Guangdong, China, and with the Medical UltraSound Image Computing (MUSIC) Lab, Shenzhen University, Shenzhen, Guangdong, China. Yan Cao and Xindi Hu are with the Shenzhen RayShape Medical Technology Inc., Shenzhen, Guangdong, China. Jin Wang and Litao Sun are with the Cancer Center, Department of Ultrasound Medicine, Zhejiang Provincial People’s Hospital, Affiliated People’s Hospital of Hangzhou Medical College, Hangzhou, Zhejiang, China. Wei Han is with the Department of Health Management Center, Qilu Hospital, Cheeloo College of Medicine, Shandong University, Jinan, Shandong, China. (email: hanwei@sdu.edu.cn; litaosun1971@sina.com; nidong@email.szu.edu.cn)}
}

\markboth{Journal of \LaTeX\ Class Files,~Vol.~14, No.~8, August~2021}%
{Shell \MakeLowercase{\textit{et al.}}: A Sample Article Using IEEEtran.cls for IEEE Journals}


\maketitle

\begin{abstract}
Three-dimensional (3D) ultrasound (US) aims to provide sonographers with the spatial relationships of anatomical structures, playing a crucial role in clinical diagnosis. 
Recently, deep-learning-based freehand 3D US has made significant advancements. It reconstructs volumes by estimating transformations between images without external tracking. 
However, image-only reconstruction poses difficulties in reducing cumulative drift and further improving reconstruction accuracy, particularly in scenarios involving complex motion trajectories.
In this context, we propose an enhanced motion network (MoNetV2) to enhance the accuracy and generalizability of reconstruction under diverse scanning velocities and tactics. 
First, we propose a sensor-based temporal and multi-branch structure that fuses image and motion information from a velocity perspective to improve image-only reconstruction accuracy.
Second, we devise an online multi-level consistency constraint that exploits the inherent consistency of scans to handle various scanning velocities and tactics.
This constraint exploits both scan-level velocity consistency, path-level appearance consistency, and patch-level motion consistency to supervise inter-frame transformation estimation. 
Third, we distill an online multi-modal self-supervised strategy that leverages the correlation between network estimation and motion information to further reduce cumulative errors.
Extensive experiments clearly demonstrate that MoNetV2 surpasses existing methods in both reconstruction quality and generalizability performance across three large datasets.
\end{abstract}


\section{Introduction}
\IEEEPARstart{U}{ltrasound} (US) imaging plays an important role in clinical monitoring and diagnosis because of its non-invasiveness, real-time, and mobility~\cite{liu2019deep}. 
Compared to 2D US, 3D US offers a more intuitive view that helps to assess the 3D shape, volume, and spatial relationships of anatomical structures more accurately. 
Its applications span various fields such as heart~\cite{demeulenaere2022coronary}, fetus~\cite{chen2023fetusmapv2}, breast~\cite{zhang2022application}, and liver~\cite{xing20223d}. 
Traditional 3D US imaging methods encompass mechanical, phased array, and freehand techniques.
Mechanical and phased array imaging often suffer from specialized and expensive hardware with a limited field of view.
In contrast, freehand imaging employs a 2D probe to reconstruct volume, affording unrestricted movement in multiple directions. 
Previous freehand methods relied on external positioning systems for high-precision positioning but encountered challenges such as high costs and susceptibility~\cite{esposito2019total}. Recently, sensorless freehand methods have made significant advances~\cite{guo2022ultrasound,luo2023recon,li2023long,yeung2024sensorless}. They mostly utilize deep learning techniques~\cite{lecun2015deep} to estimate inter-frame transformations primarily based on image content.

\begin{figure}[t]
\centerline{\includegraphics[width=\columnwidth]{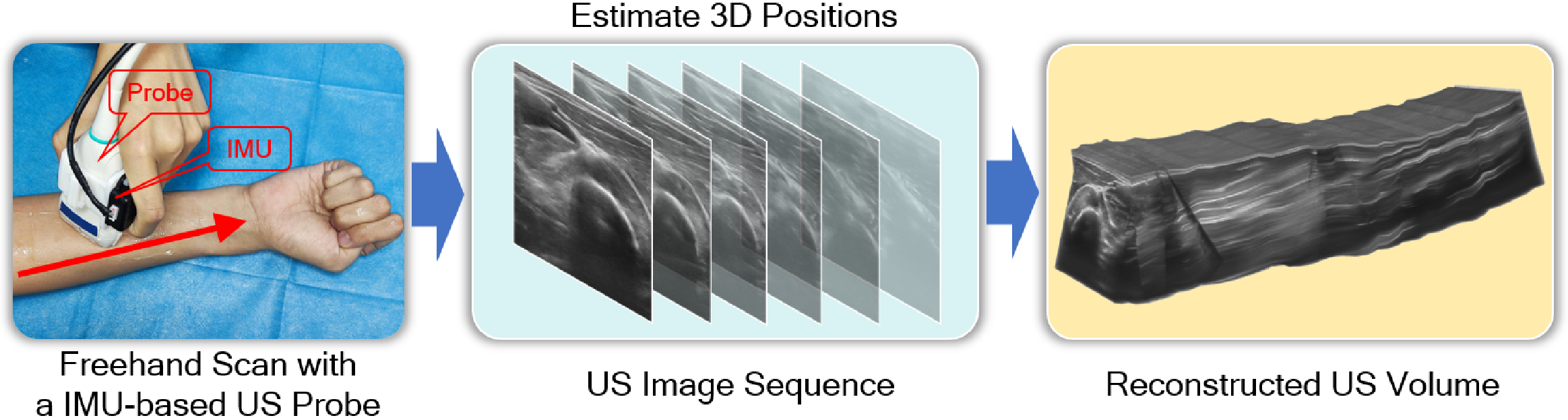}}
\caption{Pipeline of freehand 3D US reconstruction with an IMU-based US probe.}
\label{fig0}
\end{figure}

However, sensorless freehand methods still face significant challenges. 
First, the motion of the US probe typically involves various scanning velocities and tactics. Typical scanning tactics include linear, curved, sector, loop, and so on~\cite{luo2023recon}. These diverse scanning velocities and tactics result in notable variations in inter-frame content changes within the scanning sequence, presenting substantial challenges for reconstruction solutions.
Second, the relatively low signal-to-noise ratio (SNR) and contrast resolution of US images~\cite{bini2014despeckling} restrict the precision of transformation estimation when relying solely on image content. This limitation is particularly evident in the elevation direction, which is perpendicular to the image plane. The cumulative errors in estimation lead to drift phenomena, causing distortions or displacements in reconstruction. 
Motion cues beyond the image offer potential solutions to the above challenges. The lightweight inertial measurement unit (IMU) sensor can supply motion information to the US scan. The IMU is proficient in measuring physical parameters like acceleration, angular velocity, and magnetic field intensity~\cite{ahmad2013reviews}. As shown in Fig.~\ref{fig0}, it is not only cost-effective and space-efficient but also offers supplementary insights into the probe's movement. This enhancement improves the precision of transformation estimation and the quality of subsequent reconstruction.

To overcome the above-mentioned challenges, we propose a freehand 3D US reconstruction method that explores the complementary motion information, multi-level consistency, and multi-modal strategy.
Our contribution is three-fold. 
First, we propose a sensor-based temporal and multi-branch structure to seamlessly integrate image and motion information. This structure exploits the physical relationship between velocity and acceleration to enhance the precision of image-only reconstruction.
Second, we devise an online multi-level consistency constraint that employs the inherent consistency of scans to handle varying scanning velocities and tactics. 
This constraint encompasses scan-level velocity consistency, path-level appearance consistency, and patch-level motion consistency to supervise the estimation of inter-frame transformations.
Third, we distill an online multi-modal self-supervised strategy that exploits the correlation between transformation estimation and motion information for pseudo-labeling optimization to further reduce the reconstruction errors.

The rest of this paper is structured as follows. Section~\ref{sec:relatedworks} presents an overview of the related research. Section~\ref{sec:methods} provides a detailed explanation of our network and its components. Section~\ref{sec:experimentalsetup} outlines the experimental setup. Section~\ref{sec:results} delves into the analysis of the experimental results. In Section~\ref{sec:disclu}, we summarize the results, discuss the remaining challenges, and outline our future work.

\section{Related Works}
\label{sec:relatedworks}

In earlier studies, Downey et al.~\cite{downey1995vascular} were pioneers in sensorless freehand 3D US reconstruction. They achieved control over probe motion through parallel sliding or rotational scanning techniques. Subsequent research primarily relied on speckle decorrelation methods~\cite{chen1997determination,tuthill1998automated}, where in-plane and out-of-plane motion components were estimated separately using image registration and the images themselves, respectively. However, speckle decorrelation is limited to fully mature speckle patterns. To overcome this, Gee et al.~\cite{gee2006sensorless} proposed an adapted elevational decorrelation scheme that accommodates coherent scattering.
Several studies have introduced machine learning techniques. Prager et al.~\cite{prager2003sensorless} presented an alternative approach that employed linear regression on the echo-envelope intensity signal. Tetrel et al.~\cite{tetrel2016learning} utilized Gaussian process regression to estimate system errors. These studies represent early attempts at sensorless 3D US reconstruction, demonstrating the feasibility of reconstruction solely from US images.

\begin{figure*}[t]
\centerline{\includegraphics[width=\textwidth]{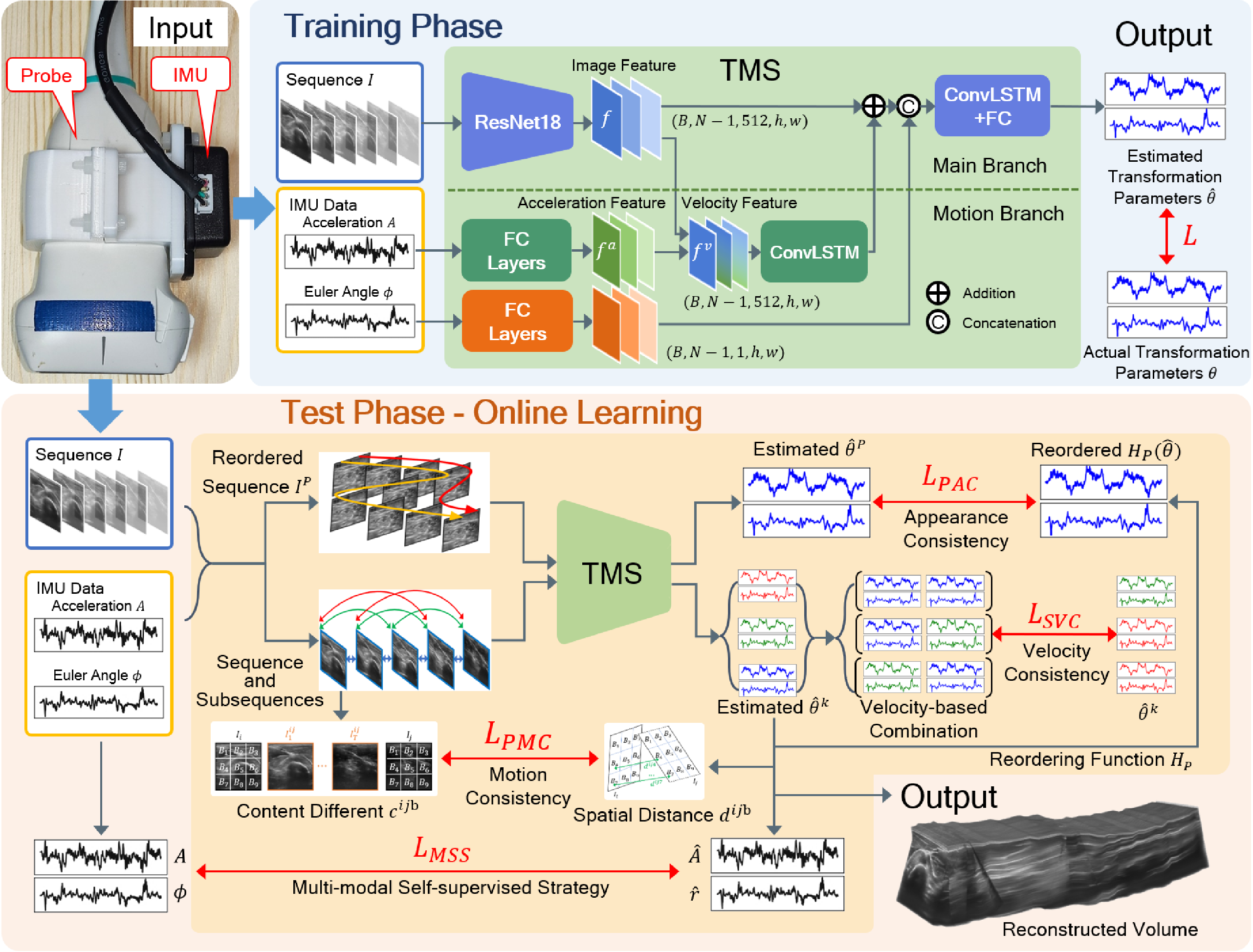}}
\caption{Pipeline of the proposed enhanced motion network (MoNetV2) for freehand 3D US reconstruction.}
\label{fig3}
\end{figure*}

Recent research has predominantly centered around deep learning~\cite{lecun2015deep}, which has demonstrated significant effectiveness in medical image tasks such as classification~\cite{hirano2021universal}, segmentation~\cite{zhang2019attention}, localization~\cite{ruan2023temporal}, diagnosis~\cite{zhang2020collaborative}, representation~\cite{sun2021context}, and anomaly detection~\cite{huang2024adapting}.
Prevost et al.~\cite{prevost2017deep} were among the first to utilize a convolutional neural network (CNN)~\cite{li2021survey} to estimate the motion of the US probe end-to-end directly from the images. 
Guo et al.~\cite{guo2020sensorless} utilized a 3D CNN to extract features between multiple US frames for smooth motion estimation. They focused the network on speckle-rich areas via a self-attention module. In their subsequent work~\cite{guo2022ultrasound}, a contrastive learning strategy was employed to leverage the label information more efficiently. 
Yeung et al.~\cite{yeung2021implicitvol,yeung2024sensorless} utilized a deep implicit representation to refine the reconstruction estimation.
Additionally, several studies explored motion and structural features within the images, such as optical flows and edges~\cite{prevost20183d,miura2020localizing,luo2023recon,kaderdina2023using,el2023trackerless}, which provide valuable information about in-plane motion and anatomical structures. To capture both temporal and global information, researchers have incorporated sequence-to-sequence models like recurrent neural networks~\cite{jordan1986serial} and transformers~\cite{vaswani2017attention,liu2023survey} in their works~\cite{luo2021self,ning2022spatial,luo2023recon,li2023trackerless,li2023long,chen2023freehand}. These models enable the capture of long-term dependencies and the context of the entire scanning sequence. Nowadays, lightweight sensors have been integrated to augment motion information beyond the US images for transformation estimation. Housden et al.~\cite{housden2008rotational} utilized orientation measurements from a micro-electro-mechanical-systems-based sensor, showcasing a substantial reduction in drift error during elevational positioning. Ito et al.~\cite{ito2017a} constructed a probe-camera system for 3D US reconstruction, employing structure from motion techniques to estimate motion. Prevost et al.~\cite{prevost20183d} integrated three-axis Euler angle from an IMU into a deep neural network, resulting in improved reconstruction accuracy. More recent studies have incorporated the acceleration from a single IMU or multiple IMUs, further exploring the potential of lightweight sensors in enhancing reconstruction performance~\cite{luo2022deep,luo2023multi}. These deep-learning-based studies have significantly reduced reconstruction errors compared to previous studies. The introduction of lightweight sensors, such as IMU, has improved reconstruction performance without significantly increasing scanning complexity. However, there are still challenges to further improving the accuracy and generalizability of 3D US reconstruction.

This study proposes an enhanced motion network (MoNetV2) to further enhance the accuracy and generalizability of 3D US reconstruction. By fusing motion information and applying multi-level consistency and multi-modal strategy, MoNetV2 can achieve more accurate inter-frame transformation estimation for varying scanning velocities and tactics. This work extends a recent conference paper~\cite{luo2022deep} with the following efforts. 
1) In MoNetV2, we propose scan-level, path-level, and patch-level consistency, which were absent in MoNet, to handle varying scanning velocities and tactics.
Scan-level velocity consistency exploits the self-prior of trajectory consistency to tackle the uneven scanning velocities.
Path-level appearance consistency leverages appearance cues to reduce cumulative drift during a loop scan.
Patch-level motion consistency exploits the fine-grained consistency between content and motion to tackle the significant variations in inter-frame content changes.
2) Compared to MoNet, we incorporate a completely new thyroid dataset (50 volunteers, 290 scans) and expand the existing arm and carotid datasets with a $25\%$ increase in volunteers, thereby broadening the scope of MoNetV2's research.
Detailed analysis reveals the higher accuracy and promising generalizability of this work.

\section{Methods}
\label{sec:methods}
\subsection{Problem Formulation}
\label{sec:methods:pf}

We equip the US probe with a lightweight IMU sensor to capture motion information, as shown in Fig.~\ref{fig0}. 
Given an $N$-length US scanning sequence as $I=\{I_i|i=1,2,\cdots,N\}$, we record the IMU orientation and acceleration corresponding to image $I_i$ as $O_i=(O_x,O_y,O_z)_i$ and $A_i=(A_x,A_y,A_z)_i$, respectively.
Here, $O_i$ represents the absolute Euler angle based on the east-north-up coordinate system. We transform the $O_i$ into the relative Euler angle $\phi_i=(\phi_x,\phi_y,\phi_z)_i$ between images $I_i$ and $I_{i+1}$ via:
\begin{equation}
\phi_i=M^{-1}\left(M(O_i)^{-1}M(O_{i+1})\right),\quad i=1,2,\cdots,N-1,
\label{eq1}
\end{equation}
where $M(\cdot)$ transforms the orientation vector into a $3\times3$ rotation matrix, and $M^{-1}(\cdot)$ represents the inverse operation of $M(\cdot)$. The gravity component of $A_i$ is denoted as $g_i$. According to the laws of physics, the $I_N$'s velocity $V_N$ can be expressed as $V_N=V_1+\sum_i (A_i-g_i)\Delta t$, where $V_1$ denotes the $I_1$'s velocity, and $\Delta t$ represents the time interval between adjacent images. Assuming that $V_1$ and $V_N$ are both equal to zero, then the mean value of $(A_i-g_i)$ is equal to zero. We update $A_i$ to minimize the influence of noise by subtracting $g_i$ from $A_i$ and adjusting the mean value of $A_i$ to zero:
\begin{equation}
A_i\leftarrow (A_i-g_i)-\frac{1}{N}\sum_{i}(A_i-g_i),\quad i=1,2,\cdots,N.
\label{eq2}
\end{equation}

Our goal is to regress the transformation parameters $\theta=\{\theta_i|i=1,2,\cdots,N-1\}$, where $\theta_i=(t_x,t_y,t_z,r_x,r_y,r_z)_i$ denotes the parameters between adjacent images $I_i$ and $I_{i+1}$. $t=\{(t_x,t_y,t_z)_i|i=1,2,\cdots,N-1\}$ and $r=\{(r_x,r_y,r_z)_i|i=1,2,\cdots,N-1\}$ represent the translation and rotation degree (Euler angle) along the three axes, respectively. Fig.~\ref{fig3} illustrates our proposed MoNetV2 framework. The input to MoNetV2 consists of all pairs of adjacent images $\{(I_i,I_{i+1})|i=1,2,\cdots,N-1\}$, the IMU acceleration $A=\{A_i|i=2,3,\cdots,N-1\}$, and the IMU Euler angle $\phi=\{\phi_i|i=1,2,\cdots,N-1\}$. MoNetV2 employs a temporal and multi-branch structure (TMS) to integrate image and motion information from a velocity perspective to regress $\theta$. 
During the test phase, two online learning strategies, incorporating an online multi-level consistency constraint (MCC) and an online multi-modal self-supervised strategy (MSS), are employed to adaptively optimize the TMS. This helps mitigate cumulative errors and improve the accuracy and generalizability of reconstruction. Further specifics will be elaborated upon in subsequent subsections.

\begin{figure*}[t]
\centerline{\includegraphics[width=\textwidth]{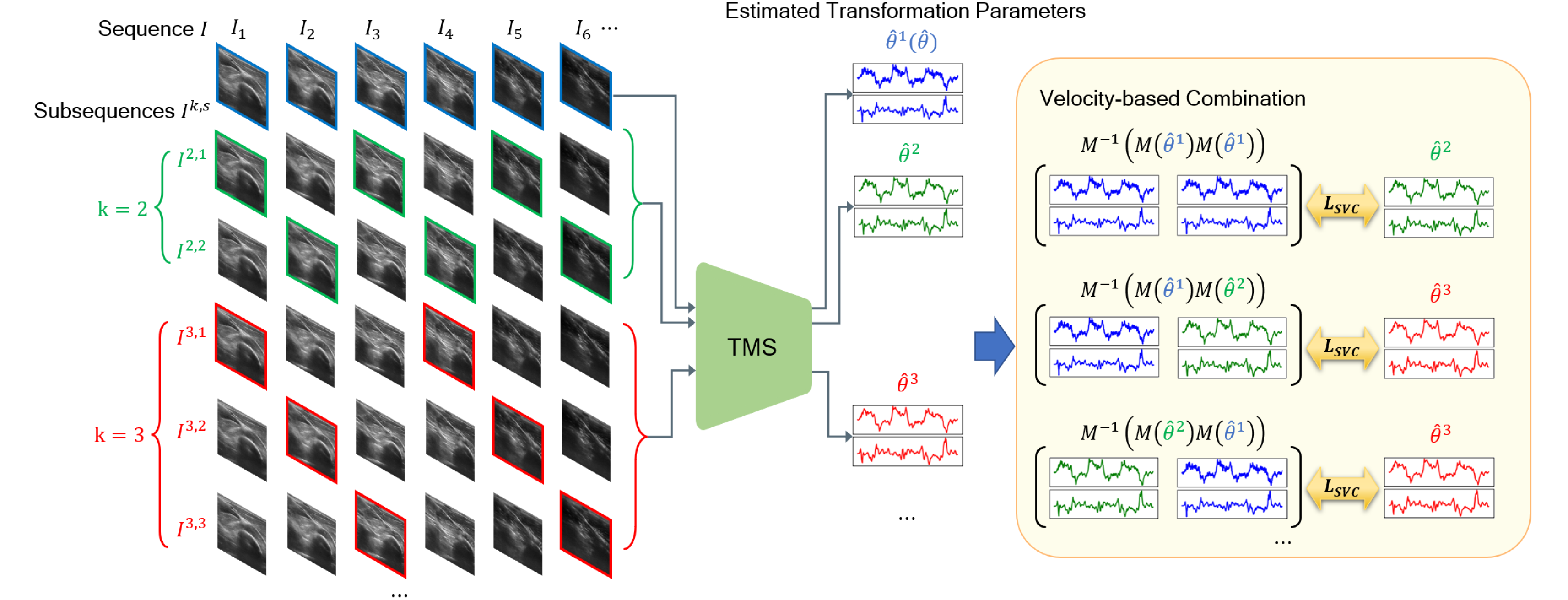}}
\caption{Overview of online scan-level velocity consistency (SVC).}
\label{fig4}
\end{figure*}

\subsection{Temporal and Multi-branch Structure}
\label{sec:methods:backbone}
Due to the significant variation in inter-frame content under different scanning velocities and tactics, relying solely on the images to estimate inter-frame transformations is challenging. We construct a temporal and multi-branch structure (TMS) to fuse image and motion information. The TMS consists of main and motion branches, as shown in the upper part of Fig.~\ref{fig3}. The main branch is a cascaded architecture of an 18-layer deep residual network (ResNet18)~\cite{he2016deep} and a single-layer convolutional long short-term memory (ConvLSTM)~\cite{shi2015convolutional}. Specifically, we removed the final fully connected layer of ResNet18 and connected it to ConvLSTM. Finally, a fully connected layer is used to generate the output. ResNet18 employs residual feature learning instead of learning features directly, resulting in improved learning efficiency and enhanced feature representation capabilities~\cite{he2020why}.  ConvLSTM preserves crucial sequential context information~\cite{greff2017lstm} that is valuable for the transformation estimation of subsequent images.
The TMS integrates IMU acceleration and Euler angle with the images, leveraging motion information beyond the images to estimate inter-frame transformation parameters more accurately.

In the main branch, ResNet18 transforms the pair of adjacent images $(I_i, I_{i+1})$ into a high-dimensional feature, denoted as $f_i$. We hold that $f_i$ implicitly contains relative distance information between the images $I_i$ and $I_{i+1}$. Assuming a consistent sampling time between adjacent images, this relative distance can be represented as the scanning velocity. In the motion branch, we associate the IMU acceleration $A$ with the high-dimensional feature $f=\{f_i|i=1,2,\cdots,N-1\}$, which contains velocity information, to achieve complementary learning and mine valuable motion cues. Specifically, we start by mapping the processed acceleration $A$ to a high-dimensional space through multiple fully connected layers. Subsequently, we reconstruct it into a 2D feature $f^a$ to facilitate integration with image features. Next, the $i$-th acceleration feature $f^a_i$ is sequentially added to the $(i-1)$-th image feature $f_{i-1}$ while retaining $f_1$, to construct the velocity feature $f^v$ from IMU acceleration:
\begin{equation}
f^v_i=
\left\{
\begin{aligned}
&f_1,\quad i=1,\\
&f_{i-1}+f^a_i,\quad i=2,3,\cdots,N-1.
\end{aligned}
\right.
\end{equation}

To minimize the impact of noise from the low-SNR acceleration $A$ on the transformation estimation, $f^v$ is fed into an ConvLSTM. This enhances the velocity feature with temporal context information, making them less susceptible to noise over a longer sequence. The output of the ConvLSTM is then integrated into the main branch. We feed the IMU Euler angle $\phi$ into multiple fully connected layers and concatenate them with the main branch's output to serve as input to the main branch's ConvLSTM. 
Since the image feature $f$ and the velocity feature $f^v$ have the same shape (see Fig.~\ref{fig3}) and both contain velocity information, adding them helps reduce feature redundancy. In contrast, the Euler angle feature has a different shape compared to the image feature, so we concatenate them instead of adding.
By integrating the IMU acceleration and Euler angle from a velocity perspective, the TMS mines valuable cues to achieve more accurate transformation estimation. 
As shown in the upper-right part of Fig.~\ref{fig3}, during the training phase, we calculate the loss between the estimated transformation parameters $\hat{\theta}$ and the ground truth $\theta$ using the mean absolute error (MAE) and Pearson correlation loss~\cite{guo2020sensorless}:
\begin{equation}
L=\left\|\hat{\theta}-\theta\right\|_1+\left(1-\frac{\textbf{Cov}(\hat{\theta},\theta)}{\sigma(\hat{\theta})\sigma(\theta)}\right),
\end{equation}
where $\left\|\cdot\right\|_1$ denotes L1 normalization, $\textbf{Cov}$ indicates the covariance, and $\sigma$ indicates the standard deviation.
The single L1 loss function applied to both translation and rotation (transformation parameters $\theta$) reflects the goal of minimizing the overall error in transformation estimation. By assigning equal weights to both components, we ensure that neither translation nor rotation is prioritized during optimization, thus preventing any bias toward one over the other.

\subsection{Online Multi-level Consistency Constraint}
\label{sec:methods:ols}
In freehand 3D US reconstruction, the variability in probe movements, encompassing different velocities and tactics, leads to shifts in data distribution. Compared to the routine strategy of offline inference, online learning offers a dynamic adjustment capability for these unseen shifts, thereby improving the precision and adaptability of reconstruction. In this study, we employ online learning to minimize cumulative drift and enhance the accuracy and generalizability of reconstruction.
As shown in the lower part of Fig.~\ref{fig3}, we propose an online multi-level consistency constraint (MCC) to effectively handle varying scanning velocities and tactics. The MCC mines the inherent consistency based on the scanning sequence itself and involves scan-, path-, and patch-level consistency.

\begin{figure*}[t]
\centerline{\includegraphics[width=\textwidth]{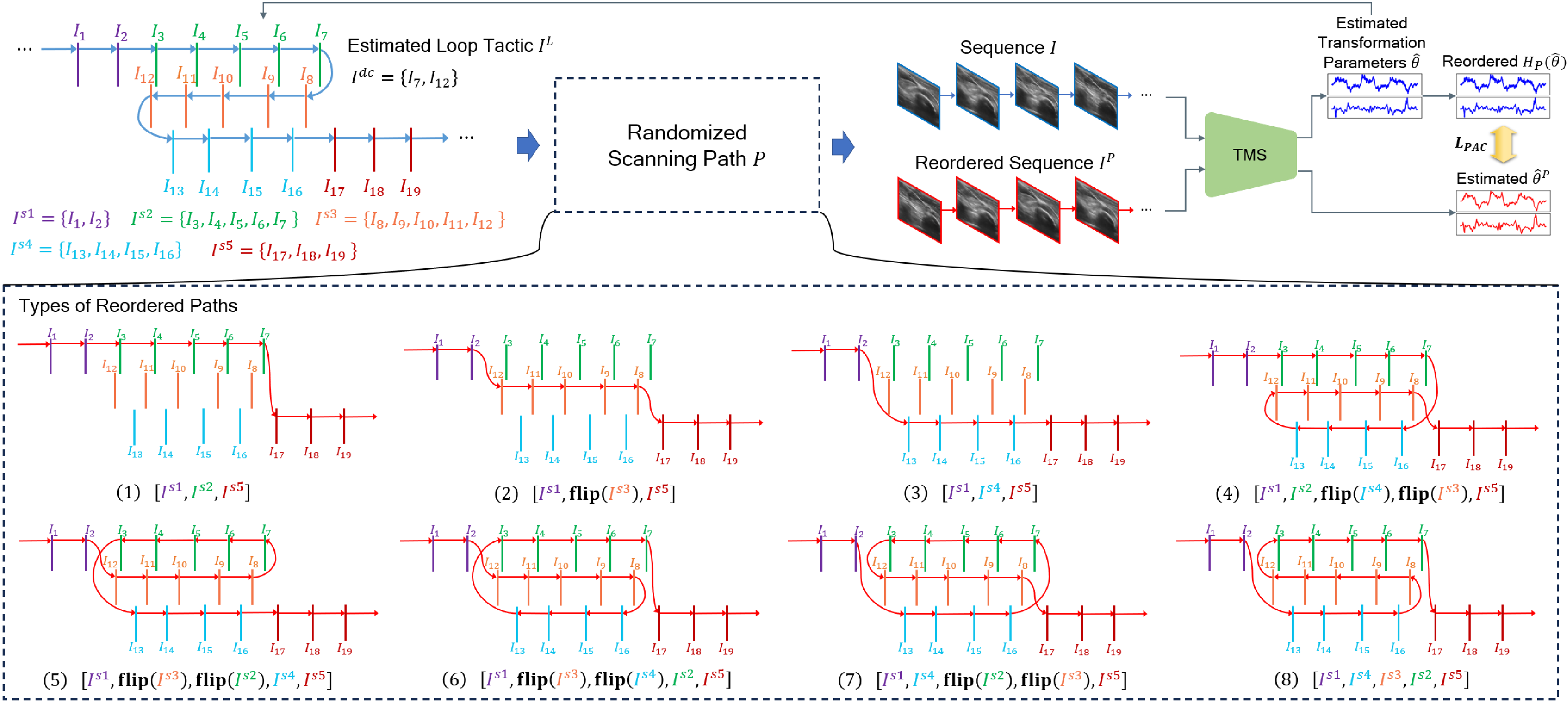}}
\caption{Overview of online path-level appearance consistency (PAC). Subsequences are distinguished by different colors.}
\label{fig_gac}
\end{figure*}

\subsubsection{Scan-level Velocity Consistency}
\label{sec:methods:ols:gsc}
Varying scanning velocities lead to significant variations in both inter-frame content changes and elevation displacements. Consequently, for scans with varying velocities but fixed trajectories, the existing models are highly likely to provide different estimated trajectories. This poses challenges for accurately estimating the inter-frame transformation. To address this challenge, we propose a scan-level velocity consistency (SVC), as illustrated in Fig.~\ref{fig4}, to exploit the self-prior of trajectory consistency  and mine velocity-independent cues between scanning sequences and their subsequences.
Given an $N$-length test scan $I$ and its corresponding IMU measurements $A$ and $\phi$, we perform interval sampling to generate subsequences to simulate various scanning velocities. These subsequences are denoted as $I^{k,s}=\{I_s,I_{s+k},I_{s+2k},\cdots\}$, where $k$ denotes interval ($2\leq k \leq K$) and $s$ indicates start index ($1\leq s\leq k$). The definitions of subsequences $A^{k,s}$ and $\phi^{k,s}$ are similar to $I^{k,s}$. In SVC, the TMS first estimates the transformation parameters $\hat{\theta}^{k,s}$ of each subsequence $I^{k,s}$ with $A^{k,s}$ and $\phi^{k,s}$. To simplify the notation, we define $\hat{\theta}^k=\bigcup_{s=1}^{k}\hat{\theta}^{k,s}=\{\hat{\theta}^k_i|i=1,2,\cdots,N-k\}$, where $\hat{\theta}^k_i$ represents the estimated transformation parameters between images $I_i$ and $I_{i+k}$. Intuitively, exact reconstruction should ensure consistency between $\hat{\theta}^{k}$ and the velocity-based combination of $\hat{\theta}^{k1}$ and $\hat{\theta}^{k-k1}$, where $1\leq k1 < k$, as shown in Fig.~\ref{fig4}. This reflects the trajectory consistency of the scans. Thus, we set a velocity consistency loss $L_{SVC}$ to minimize the MAE between $\hat{\theta}^k$ and the combination of $\hat{\theta}^{k1}$ and $\hat{\theta}^{k-k1}$.
\begin{equation}
L_{SVC}=\frac{1}{Z}\sum^K_{k=2}\sum^{k-1}_{k1=1}\sum^{N-k}_{i=1}\left|\hat{\theta}^k_i-M^{-1}\left(M(\hat{\theta}^{k1}_i)M(\hat{\theta}^{k-k1}_{i+k1})\right)\right|,
\end{equation}
where $Z=\sum^K_{k=2}\sum^{k-1}_{k1=1}1$ is the average factor, $M(\cdot)$ transforms the transformation parameters into a $4\times 4$ transformation matrix, and $M^{-1}(\cdot)$ represents the inverse operation of $M(\cdot)$. This enhances the TMS generalizability against the diverse scanning velocities or frame rates.

\subsubsection{Path-level Appearance Consistency}
\label{sec:methods:ols:gca}
The scanning probe may revisit areas already scanned when changing direction. This scenario results in sequentially distant yet spatially close images that share contextual similarity. 
The appearance consistency of the anatomical structures ensures that these similar images should have a close estimated position.
However, this is challenging due to the cumulative drift of the sequence.
To tackle this challenge, we propose a path-level appearance consistency (PAC), as shown in Fig.~\ref{fig_gac}, to exploit appearance cues and reduce cumulative drift.
In PAC, we utilize the TMS to estimate the transformation parameters $\hat{\theta}$ for a test scan $I$ with corresponding IMU data.
For each estimated loop tactic $I^L$ in sequence $I$, we first identify the frames with direction changes, denoted as $I^{dc}$. We then divide $I^L$ into several subsequences ($I^{s1}$, $I^{s2}$, $I^{s3}$, $I^{s4}$, and $I^{s5}$ in Fig.~\ref{fig_gac}) with unchanged scanning direction based on $I^{dc}$. We randomly select a scanning path $P$ from all subsequences in scan $I$ to create a reordered sequence, denoted as $I^{P}$. The reordered paths for each loop tactic in $P$ include the following 8 types:
\begin{enumerate}
\item $[I^{s1},I^{s2},I^{s5}]$,
\item $[I^{s1},\textbf{flip}(I^{s3}),I^{s5}]$,
\item $[I^{s1},I^{s4},I^{s5}]$,
\item $[I^{s1},I^{s2},\textbf{flip}(I^{s4}),\textbf{flip}(I^{s3}),I^{s5}]$,
\item $[I^{s1},\textbf{flip}(I^{s3}),\textbf{flip}(I^{s2}),I^{s4},I^{s5}]$,
\item $[I^{s1},\textbf{flip}(I^{s3}),\textbf{flip}(I^{s4}),I^{s2},I^{s5}]$,
\item $[I^{s1},I^{s4},\textbf{flip}(I^{s2}),\textbf{flip}(I^{s3}),I^{s5}]$,
\item $[I^{s1},I^{s4},I^{s3},I^{s2},I^{s5}]$,
\end{enumerate}
where $\textbf{flip}(\cdot)$ denotes the flip operation of the sequence. Subsequently, the TMS estimates the transformation parameters $\hat{\theta}^P$ for sequences $I^P$ with corresponding IMU data. Intuitively, $\hat{\theta}^P$ should be consistent with $\hat{\theta}$ that underwent the same reordering, to reflect 3D appearance consistency. Therefore, we set a appearance consistency loss $L_{PAC}$ to $\hat{\theta}^P$ and the reordered $\hat{\theta}$.
\begin{equation}
L_{PAC}=\left\|\hat{\theta}^P-H_P(\hat{\theta})\right\|_1,
\end{equation}
where $H_P$ reorders the transformation parameters under the selected scanning path $P$.

\begin{figure}[t]
\centerline{\includegraphics[width=\columnwidth]{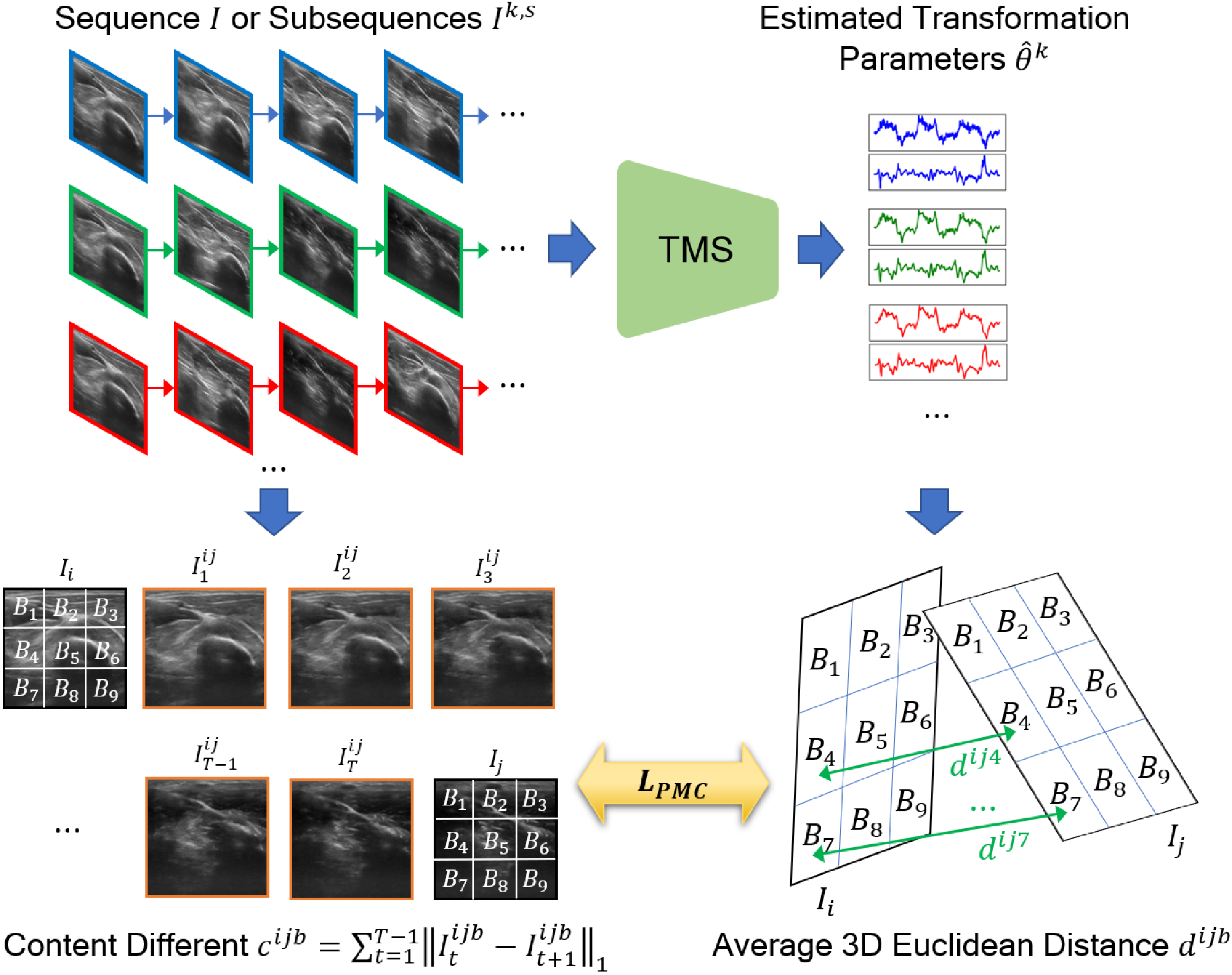}}
\caption{Overview of online patch-level motion consistency (PMC). $B_{b}$ denotes the $b$-th patch.}
\label{fig6}
\end{figure}

\subsubsection{Patch-level Motion Consistency}
\label{sec:methods:ols:lis}
The diverse tactics result in significant variations in inter-frame content changes in a scanning sequence. For instance, in sector or curved tactics, neighboring frames even exhibit inconsistent content changes at different vertical or horizontal positions inside the frame. This inconsistency highlights the complexity of inter-frame motion, posing challenges to accurately estimating the inter-frame transformation. To address this challenge, we propose a patch-level motion consistency (PMC), as illustrated in Fig.~\ref{fig6}, to exploit the correlation between content and motion in fine-grained image patches.
In PMC, for an $N$-length test scan $I$ and its subsequences $I^{k,s}$ (see subsection~\ref{sec:methods:ols:gsc}), we first employ the FILM model~\cite{reda2022film} to interpolate any pair of adjacent images $(I_i,I_j)$, generating $T$ images $G^{ij}=\{I^{ij}_1,I^{ij}_2,\cdots,I^{ij}_T\}$ in the image space between $I_i$ and $I_j$. $G^{ij}$ can be considered as an approximate geodesic between the images $I_i$ and $I_j$, which captures complex nonlinear changes between two adjacent images, making the content differences more accurate and physically meaningful than directly using two adjacent images. As shown in Fig.~\ref{fig6}, we then divide each image into $P\times Q$ (we set $32\times 32$) patches, where $P$ and $Q$ determine the numbers of rows and columns of patches, respectively. We quantify the content difference of the $b$-th patch between $I_i$ and $I_j$ as $c^{ijb}$ using the geodetic distance, which sums the patch content differences between adjacent images on $G^{ij}$.
\begin{equation}
c^{ijb}=\sum^{T-1}_{t=1}\left\|I^{ijb}_t-I^{ijb}_{t+1}\right\|_1,
\end{equation}
where $I^{ijb}_t$ represents the $b$-th patch of $I^{ij}_t\in G^{ij}$, and $1\leq b \leq P\times Q$. Meanwhile, after estimating the transformation parameters using the TMS, we calculate the average 3D Euclidean distance $d^{ijb}$ of the $b$-th patch between the images $I_i$ and $I_j$ to measure the fine-grained probe motion.
Intuitively, image patch with significant content difference should exhibit more substantial motion, implying relatively larger average Euclidean distance. 
It should be noted that the average Euclidean distances from different patches reflect the relative rotation.
This reflects the local content and motion consistency of the scans. Therefore, we set a fine-grained motion consistency loss $L_{PMC}$ to minimize the MAE between the normalized content difference and the normalized distance.
\begin{equation}
L_{PMC}=\frac{1}{Z}\sum_{k=1}^K\sum_{\substack{1\leq i,j \leq N\\j=i+k}}\sum^{P\times Q}_{b=1}\left\|\frac{c^{ijb}-\mu(c)}{\sigma(c)}-\frac{d^{ijb}-\mu(d)}{\sigma(d)}\right\|_1,
\end{equation}
where $Z=\sum_{k=1}^K\sum_{1\leq i,j \leq N,j=i+k}\sum^{P\times Q}_{b=1}1$ denotes the average factor, $c=\{c^{ijb}|1\leq i,j \leq N,j=i+k,1\leq k \leq K,1\leq b\leq P\times Q\}$, $d=\{d^{ijb}|1\leq i,j \leq N,j=i+k,1\leq k \leq K,1\leq b\leq P\times Q\}$, $\mu$ calculates the mean value, and $\sigma$ calculates the standard deviation.

\subsection{Online Multi-modal Self-supervised Strategy}
\label{sec:methods:ols:imu}

We propose an online multi-modal self-supervised strategy (MSS) that leverages the correlation between the TMS estimation and the IMU measurements to further reduce accumulated drift. As shown in the lower part of Fig.~\ref{fig3}, after estimating the inter-frame transformation parameters $\hat{\theta}$ using the TMS, we use the IMU acceleration and Euler angle as pseudo-labels for adaptive optimization. We calculate the estimated acceleration $\hat{A}$ at the center point of each image based on the estimated translation $\hat{t}$. To mitigate the impact of noise, we scale the calculated acceleration to match the scale of IMU acceleration with a zero mean:
\begin{equation}
\hat{A}_i=\hat{t}^{-1}_{i-1}+\hat{t}_{i}-\frac{1}{N-2}\sum_{i}\left(\hat{t}^{-1}_{i-1}+\hat{t}_{i}\right),\quad i=2,\cdots,N-1,
\end{equation}
where $\hat{t}^{-1}_{i-1}$ denotes the translation in the inversion of $\hat{\theta}_{i-1}$.
The multi-modal self-supervised loss $L_{MSS}$ comprises two components: acceleration and the Euler angle. We use the Pearson correlation loss to quantify the disparity between the estimated acceleration $\hat{A}$ and the IMU acceleration $A$, aiming to enhance the TMS estimation by capturing more accurate large-scale trends. For the estimated Euler angle $\hat{r}$ and the IMU Euler angle $\phi$, we apply the MAE loss as a constraint.
\begin{equation}
L_{MSS}=\left(1-\frac{\textbf{Cov}(\hat{A},A)}{\sigma(\hat{A})\sigma(A)}\right)+\left\|\hat{r}-\phi\right\|_1.
\end{equation}

The four online losses ($L_{SVC}$, $L_{PAC}$, $L_{PMC}$, and $L_{MSS}$) are equally important, and their numerical differences are not large. Therefore, we set the weights of all these losses to 1, ensuring that MoNetV2 treats each loss equally without favoring any one over the others.

\section{Experimental Setup}
\label{sec:experimentalsetup}

\subsection{Datasets}
\label{sec:experimentalsetup:datasets}

\begin{figure}[t]
\centerline{\includegraphics[width=\columnwidth]{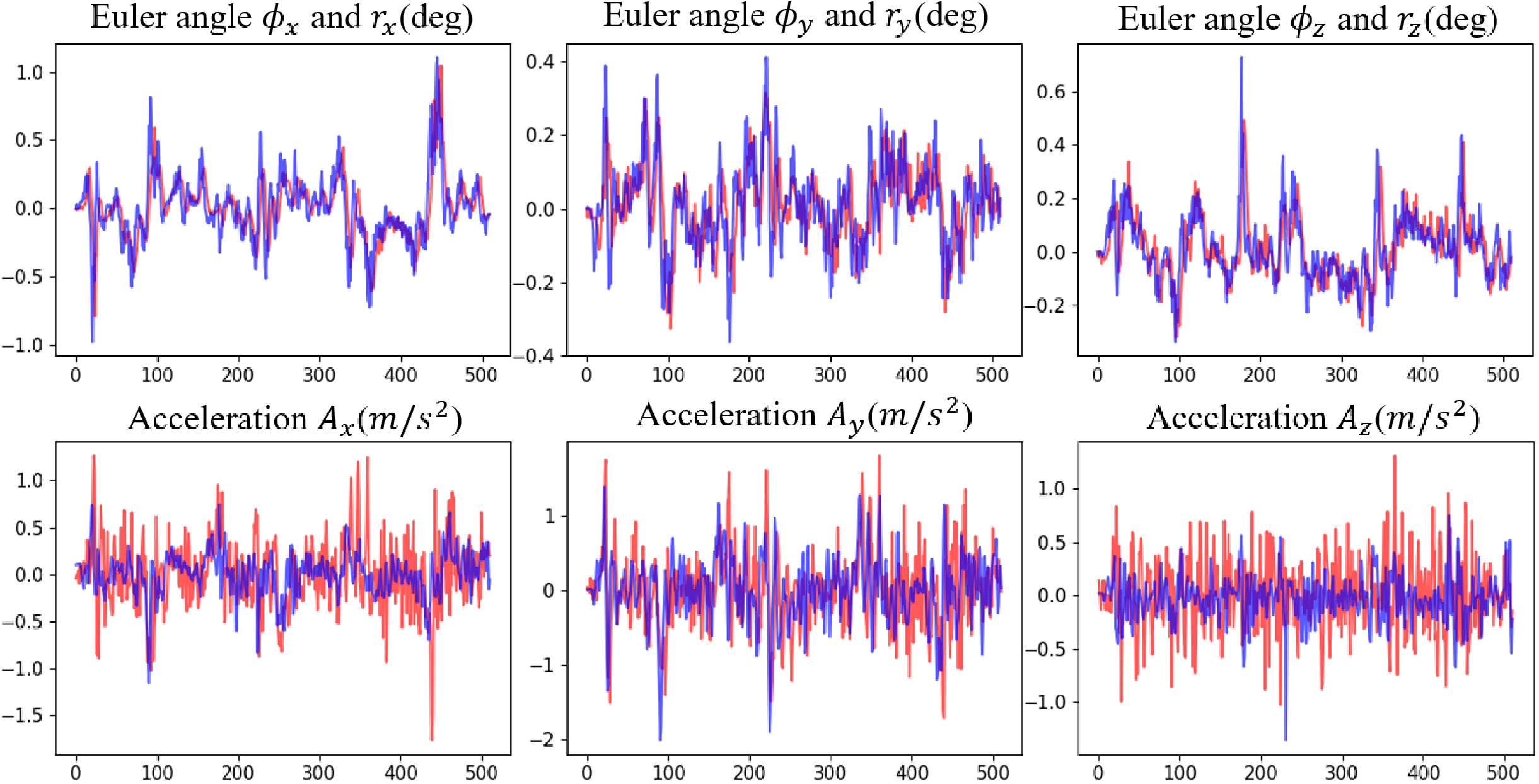}}
\caption{Comparison of IMU data ($\phi$ and $A$, blue line) and EM positioning data ($r$ and $A$, red line). The abscissa of each subfigure indicates the image index.}
\label{fig8}
\end{figure}

We established a data collection platform to collect all US scans and corresponding IMU data. The platform primarily comprises a portable US machine (M8, Shenzhen Mindray Bio-Medical Electronics Co., Ltd., China), a linear probe (L12-4s, Shenzhen Mindray Bio-Medical Electronics Co., Ltd., China), an IMU sensor (WT901C-232, WitMotion ShenZhen Co., Ltd., China), and an electromagnetic (EM) positioning system (3D Guidance trakSTAR, Northern Digital Inc., Canada), which includes an electronics unit, transmitter, and sensor. As shown in Fig.~\ref{fig0}, we used a 3D-printed adapter to bind the IMU to the US probe, enabling us to obtain acceleration and orientation information while scanning US images. We simultaneously bound the EM sensor to the US probe to precisely track the scanning trajectory using EM positioning. The measurement resolutions for IMU acceleration and orientation were $5\times 10^{-4}$ g/LSB and $0.5$ degree, respectively. The EM position and orientation resolution were $1.4$ mm and $0.5$ degree, respectively. We calibrated the data collection platform using the Levenberg-Marquardt algorithm~\cite{levenberg1944method}, ensuring the accuracy of IMU and EM measurements while minimizing system errors. For each frame, we simultaneously acquire ultrasound images, EM data, and IMU data at a frequency of 30 fps to ensure alignment. The ground truth of the inter-frame transformation parameters $\theta$ is calculated from the absolute transformation parameters measured by the EM positioning system. We compared the preprocessed IMU data (see Eqs.~\ref{eq1} and~\ref{eq2}) with the EM positioning data, as shown in Fig.~\ref{fig8}. The IMU Euler angle closely corresponds to the EM rotation degree. In contrast, the IMU acceleration exhibits significant noise at single points but follows a consistent trend with EM data over a broader range.

We utilized the data collection platform to construct three datasets involving 50 volunteers, including the arm, carotid, and thyroid. The arm dataset comprises 583 scans, with an average length of $386.39$ mm. Scanning tactics included linear, curved, loop, and sector scans, which simulate clinical scenarios. The carotid dataset contains 432 scans with an average length of $241.25$ mm, including linear, loop, and sector scanning tactics. The thyroid dataset contains 290 scans with an average length of $66.25$ mm, including linear and loop scanning tactics. The depth of all US images is $4$ cm. The image size is $248\times 260$ pixels, and the image spacing is $0.15\times 0.15$ mm$^2$. The collection and use of the data received approval from the local institutional review board (IRB).

\subsection{Evaluation Metrics}
\label{sec:experimentalsetup:metrics}
For an $N$-length scan $I$, we calculate the actual spatial position $P=\{P_i|i=1,2,\cdots,N\}$ based on the ground truth of transformation parameters $\theta$, where each $P_i$ denotes the spatial coordinates corresponding to the center point of frame $I_i$. The calculation can be expressed as
\begin{equation}
\left[
\begin{matrix}
\mathbf{R} & P_i \\
0 & 1
\end{matrix}
\right]=
\left\{
\begin{aligned}
&\mathbf{I},\quad i=1,\\
&\prod^i_{k=2}M(\theta_{k-1}),\quad i=2,3,\cdots,N,
\end{aligned}
\right.
\end{equation}
where $M(\cdot)$ transforms the transformation parameters into a $4\times 4$ transformation matrix, $\mathbf{I}$ denotes a $4\times 4$ identity matrix, and $\mathbf{R}$ denotes a $3\times 3$ rotation matrix. 
Similarly, we calculate the estimated spatial position $\hat{P}=\{\hat{P}_i|i=1,2,\cdots,N\}$ based on the estimated transformation parameters $\hat{\theta}$.
We evaluate the reconstruction performance using the following six metrics consistent with Luo et al.~\cite{luo2023recon}. 
For all metrics, lower values indicate better reconstruction performance.
\begin{enumerate}
\item \textbf{Final Drift Rate (FDR)}, the ratio of the distance between the estimated and the actual positions (drift) of the last frame of the scan to the scan length:
\begin{equation}
\textit{FDR}=\frac{\|\hat{P}_N-P_N\|_2}{\sum^{N-1}_{k=1}\|P_{k+1}-P_k\|_2},
\end{equation}
where $\|\cdot\|_2$ denotes L2 normalization.
\item \textbf{Average Drift Rate (ADR)}, the mean value of the drift rate, where the drift rate is calculated as the ratio of the cumulative drift of a frame to the length from that frame to the initial frame of the scan:
\begin{equation}
\textit{ADR}=\frac{1}{N-1}\sum^N_{i=2}\frac{\|\hat{P}_i-P_i\|_2}{\sum^{i-1}_{k=1}\|P_{k+1}-P_k\|_2}.
\end{equation}
\item \textbf{Maximum Drift (MD)}, the maximum drift among all frames:
\begin{equation}
\textit{MD}=\max_{\hat{P}_i\in \hat{P},P_i\in P}\|\hat{P}_i-P_i\|_2.
\end{equation}
\item \textbf{Sum of Drift (SD)}, the total sum of all drifts:
\begin{equation}
\textit{SD}=\sum^N_{i=1}\|\hat{P}_i-P_i\|_2.
\end{equation}
\item \textbf{Symmetric Hausdorff Distance (HD)}, the maximum distance between the estimated and the actual positions across all frames:
\begin{equation}
\begin{split}
\textit{HD}=\max\bigg(&\max_{\hat{P}_i\in \hat{P}}\min_{P_i\in P}\|\hat{P}_i-P_i\|_2,\\
&\max_{P_i\in P}\min_{\hat{P}_i\in \hat{P}}\|P_i-\hat{P}_i\|_2\bigg).
\end{split}
\end{equation}
\item \textbf{Mean Error of Angle (MEA)}, the MAE between the estimated Euler angle $\hat{r}$ and the actual Euler angle $r$:
\begin{equation}
\textit{MEA}=\left\|\hat{r}-r\right\|_1.
\end{equation}
\end{enumerate}

\begin{table*}[t]
\caption{Mean(std) results of the reconstruction performance of different methods. TMS/I: our TMS with images as input. The best results are shown in blue.}
\centering
\begin{tabular}{p{25pt}|p{67pt}|p{50pt}|p{50pt}|p{50pt}|p{70pt}|p{50pt}|p{45pt}}
\hline\hline
\multicolumn{1}{c|}{Dataset}& \multicolumn{1}{|c|}{Method}& \multicolumn{1}{|c|}{FDR(\%)$\downarrow$}& \multicolumn{1}{|c|}{ADR(\%)$\downarrow$}& \multicolumn{1}{|c|}{MD(mm)$\downarrow$}& \multicolumn{1}{|c|}{SD(mm)$\downarrow$}& \multicolumn{1}{|c|}{HD(mm)$\downarrow$}& \multicolumn{1}{|c}{MEA(deg)$\downarrow$} \\
\hline
\multirow{8}*{Arm}& CNN-OF& $33.03(21.26)$& $46.80(36.51)$& $82.94(36.87)$& $2360.97(1503.33)$& $74.85(36.37)$& $4.82(3.02)$ \\
~& ResNet18& $21.13(12.75)$& $31.29(18.03)$& $55.01(28.18)$& $1684.58(1795.82)$& $51.10(26.91)$& $7.36(4.19)$ \\
~& DC$^2$-Net& $18.12(12.70)$& $26.63(15.03)$& $48.26(30.17)$& $1399.03(1607.15)$& $48.15(30.76)$& $7.02(3.99)$ \\
~& RecON& $15.37(9.65)$& $22.23(11.67)$& $34.41(20.11)$& $1096.15(962.99)$& $30.97(15.75)$& $5.23(3.46)$ \\
~& MoNet& $14.38(8.67)$& $21.20(10.44)$& $32.36(18.71)$& $1009.60(863.52)$& $28.96(14.17)$& $3.70(2.30)$ \\
~& OSCNet& $13.06(7.41)$& $19.90(11.22)$& $30.81(17.29)$& $947.06(716.63)$& $27.69(13.21)$& $3.45(2.22)$ \\
\cline{2-8}
~& TMS/I& $18.30(12.29)$& $27.07(17.22)$& $46.87(32.33)$& $1586.31(1932.69)$& $44.13(31.09)$& $6.28(4.30)$ \\
~& MoNetV2& \textcolor{blue}{$\mathbf{11.04(6.02)}$}& \textcolor{blue}{$\mathbf{18.26(9.10)}$}& \textcolor{blue}{$\mathbf{28.84(15.62)}$}& \textcolor{blue}{$\mathbf{901.87(658.43)}$}& \textcolor{blue}{$\mathbf{26.24(12.47)}$}& \textcolor{blue}{$\mathbf{3.00(2.25)}$} \\
\hline
\multirow{8}*{Carotid}& CNN-OF& $28.25(18.25)$& $42.87(21.57)$& $45.12(17.47)$& $1392.58(1056.99)$& $39.68(16.56)$& $3.95(2.91)$ \\
~& ResNet18& $21.47(13.47)$& $32.56(13.05)$& $37.53(16.90)$& $1157.17(740.42)$& $33.24(15.61)$& $5.34(3.23)$ \\
~& DC$^2$-Net& $19.06(13.00)$& $30.64(17.08)$& $33.06(15.15)$& $1017.02(814.82)$& $27.99(12.81)$& $5.43(3.15)$ \\
~& RecON& $15.74(10.45)$& $26.80(18.99)$& $24.90(11.22)$& $800.50(716.87)$& $22.36(11.26)$& $4.25(2.78)$ \\
~& MoNet& $14.53(9.51)$& $26.50(19.23)$& $23.67(10.69)$& $753.40(593.38)$& $21.11(10.95)$& $2.92(1.76)$ \\
~& OSCNet& $14.17(9.46)$& $25.42(19.00)$& $23.25(10.53)$& $714.22(526.48)$& $20.62(10.54)$& $2.69(1.67)$ \\
\cline{2-8}
~& TMS/I& $19.88(17.40)$& $32.21(21.34)$& $32.20(14.94)$& $898.44(606.06)$& $27.20(11.90)$& $4.55(2.20)$ \\
~& MoNetV2& \textcolor{blue}{$\mathbf{12.11(7.55)}$}& \textcolor{blue}{$\mathbf{23.78(17.51)}$}& \textcolor{blue}{$\mathbf{20.75(8.82)}$}& \textcolor{blue}{$\mathbf{646.00(455.03)}$}& \textcolor{blue}{$\mathbf{18.50(8.64)}$}& \textcolor{blue}{$\mathbf{2.40(1.55)}$} \\
\hline
\multirow{8}*{Thyroid}& CNN-OF& $28.82(12.75)$& $47.01(23.14)$& $13.79(10.83)$& $419.98(377.58)$& $12.35(10.42)$& $1.86(0.92)$ \\
~& ResNet18& $26.95(12.38)$& $43.88(25.68)$& $13.04(9.70)$& $397.94(369.21)$& $12.85(9.12)$& $2.69(1.21)$ \\
~& DC$^2$-Net& $25.76(11.49)$& $37.25(23.97)$& $12.71(9.45)$& $302.97(353.10)$& $11.80(8.79)$& $2.49(2.09)$ \\
~& RecON& $22.65(10.12)$& $31.15(15.40)$& $10.48(5.84)$& $239.92(222.34)$& $10.11(7.47)$& $1.95(1.06)$ \\
~& MoNet& $19.64(9.38)$& $29.17(11.36)$& $10.16(5.87)$& $228.12(210.99)$& $9.63(5.82)$& $1.56(1.16)$ \\
~& OSCNet& $18.78(9.07)$& $28.44(10.91)$& $9.36(5.11)$& $224.83(211.32)$& $8.68(6.23)$& $1.50(1.14)$ \\
\cline{2-8}
~& TMS/I& $24.01(13.99)$& $36.84(18.68)$& $12.01(9.46)$& $327.58(369.98)$& $10.49(7.55)$& $2.12(1.25)$ \\
~& MoNetV2& \textcolor{blue}{$\mathbf{17.88(9.10)}$}& \textcolor{blue}{$\mathbf{26.34(9.12)}$}& \textcolor{blue}{$\mathbf{8.93(4.79)}$}& \textcolor{blue}{$\mathbf{207.60(195.20)}$}& \textcolor{blue}{$\mathbf{8.43(4.73)}$}& \textcolor{blue}{$\mathbf{1.36(0.97)}$} \\
\hline\hline
\end{tabular}
\label{tab1}
\end{table*}

\begin{figure}[t]
\centerline{\includegraphics[width=\columnwidth]{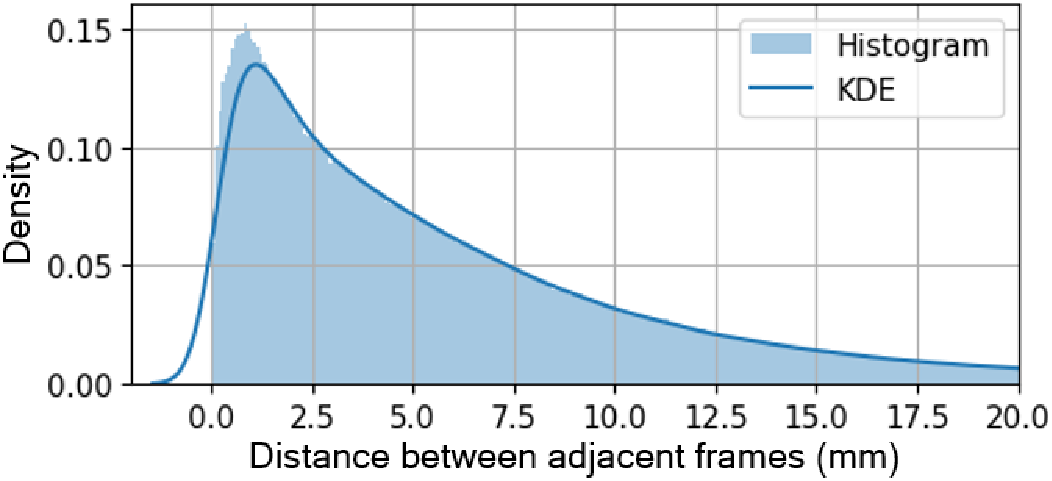}}
\caption{The histogram and kernel density estimation (KDE) of the distance between adjacent frames.}
\label{fig:distance}
\end{figure} 

\subsection{Experimental Settings}
\label{sec:experimentalsetup:setip}
We randomized the arm, carotid, and thyroid datasets into 375/104/104, 276/78/78, and 184/53/53 scans based on volunteer level to construct the training/validation/test sets, respectively. To enhance the robustness of the network and mitigate overfitting risks, we performed random augmentations on each scan, including subsequence intercepting, interval sampling, and sequence inversion. In the training set, we augmented each scan into 20 different sequences, regenerating these augmented sequences in each training epoch to simulate a complex and diverse range of scanning scenarios. In the validation and test sets, we fixed each scan to ten augmented sequences to ensure consistency in testing. For the arm and carotid datasets, the sampling interval range was limited to 1 to 11 frames. For the thyroid dataset, the sampling interval range was limited to 1 to 4 frames due to its relatively short sequence. Fig.~\ref{fig:distance} shows the distribution of the distance between the adjacent frames after sampling. The sampled inter-frame distances cover a range from short to long distances, consistent with real clinical scanning situations.
We used the Adam optimizer~\cite{kingma2014adam} to optimize MoNetV2. During the training phase, the epoch was set to 200, and the batch size was 1. To prevent overfitting, we initialized the learning rate at $10^{-4}$ and employed a learning rate decay strategy, reducing the learning rate by half every 30 epochs. During the test phase, the number of iterations and learning rate for online learning were set to 60 and $2\times10^{-6}$, respectively. All code was implemented in PyTorch~\cite{paszke2019pytorch} and ran on a server equipped with RTX 3090 GPUs.

\section{Experiments and Results}
\label{sec:results}

\subsection{Comparison of Different Methods}
\label{sec:results:diffmethods}
To validate the effectiveness of the proposed MoNetV2, we conducted a comparison with five state-of-the-art 3D US reconstruction methods: CNN-OF~\cite{prevost20183d}, DC$^2$-Net~\cite{guo2022ultrasound}, RecON~\cite{luo2023recon}, MoNet~\cite{luo2022deep}, and OSCNet~\cite{luo2023multi}. We also compared MoNetV2 with the well-known ResNet18~\cite{he2016deep} in the field of computer vision.
CNN-OF is a pioneer in leveraging deep learning to estimate the relative transformation of US images. It uses a pair of adjacent frames, optical flow, and IMU Euler angle as input to estimate inter-frame transformation parameters.
DC$^2$-Net introduces the attention mechanism and contrast learning to optimize relative position estimation by focusing on speckle-rich regions and mining label information. It takes multiple consecutive frames as input and estimates their average transformation parameters. However, parameter averaging is not suitable for trajectories with various scanning tactics, as it aims to smooth jitter in scans with slight variations. In our comparison, DC$^2$-Net takes only a pair of adjacent frames as input.
RecON pioneers research on 3D US reconstruction under intricate scanning tactics. It effectively addresses diverse scanning velocities and poses stemming from complex tactics by utilizing motion-weighted loss, online self-supervised learning, and differentiable reconstruction approximation. In our comparison, we removed RecON's shape prior module due to the lack of distinct shape features in muscles and vessels. RecON takes the scanning sequence, optical flow, and canny edge map as input to estimate all transformation parameters of the scanning sequence.
MoNet introduces IMU information and online multi-modal strategy to estimate transformation parameters. It takes the scanning sequence and IMU data as input.
OSCNet introduces online self-supervised strategies based on US images and multiple IMUs to improve transformation estimation. It takes the scanning sequence and multiple IMU data as input. As the data we collected is derived from a single IMU, OSCNet takes only single IMU data as input to estimate all transformation parameters of the scanning sequence and removes multi-IMU losses in our comparison.
ResNet18 takes a pair of adjacent frames as input to estimate their transformation parameters.

\begin{figure*}[t]
\centerline{\includegraphics[width=\textwidth]{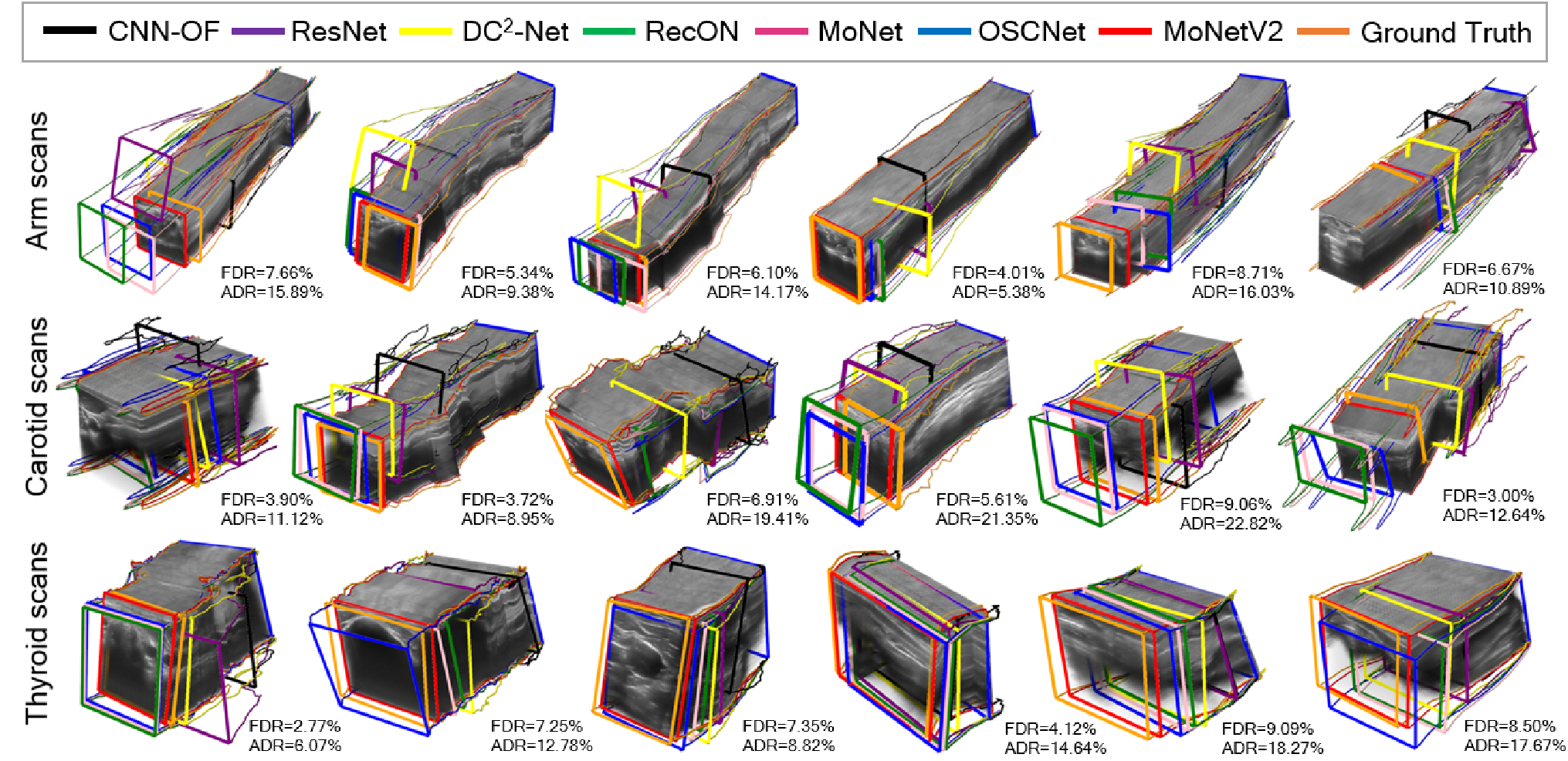}}
\caption{Typical reconstruction cases on arm (Row I), carotid (Row II), and thyroid (Row III) scans. The FDR and ADR values of MoNetV2 are shown in the lower right of each case.}
\label{fig9}
\end{figure*}

\begin{figure}[t]
\centerline{\includegraphics[width=\columnwidth]{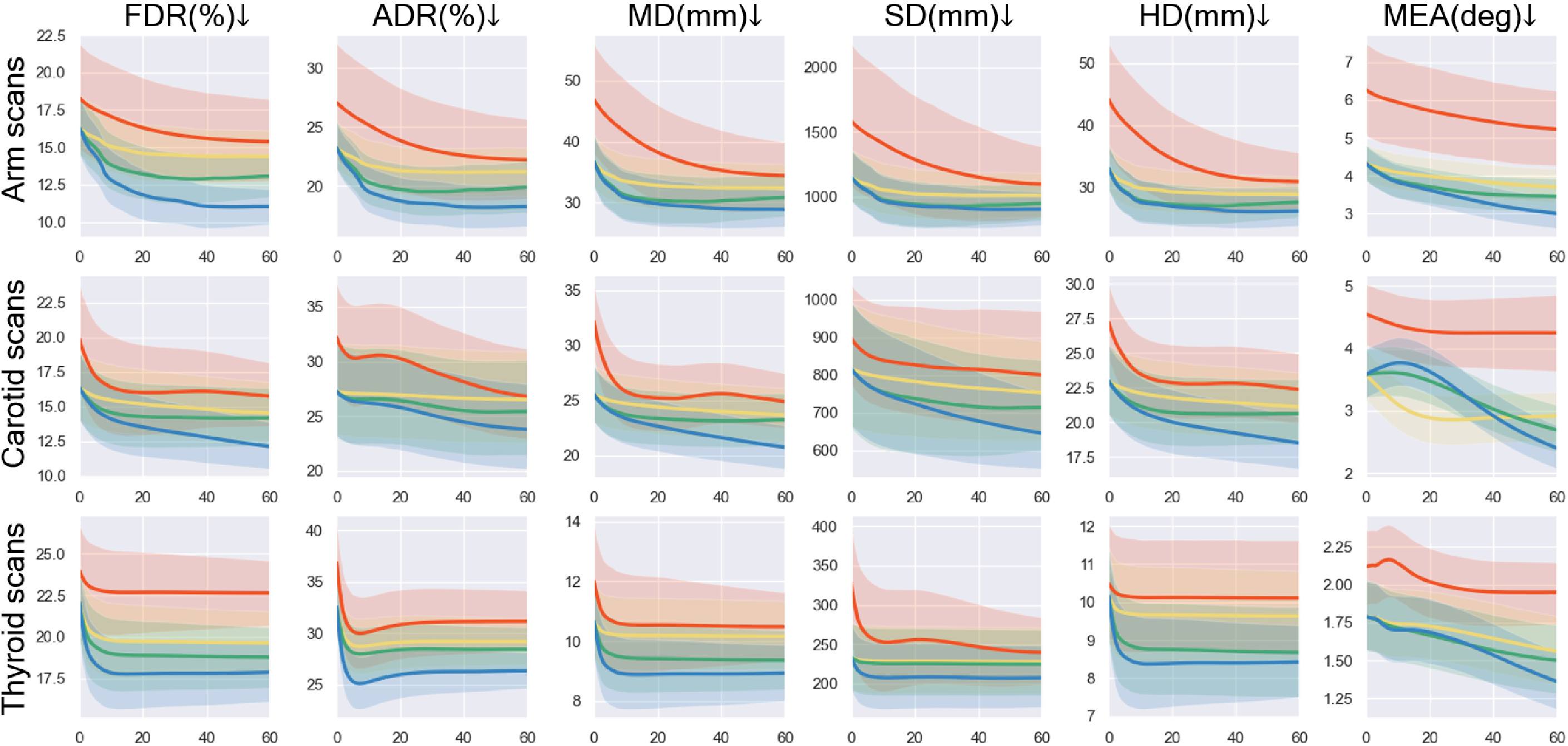}}
\caption{Metric curves (with 95\% confidence interval) declines for the online optimization on arm/carotid/thyroid (Row I/II/III) scans. Red: RecON, Yellow: MoNet, Green: OSCNet, and Blue: MoNetV2. The x- and y-axis show the number of iterations and metrics, respectively.}
\label{fig10}
\end{figure}

\begin{table*}[t]
\caption{Mean(std) results of the reconstruction performance on different inputs and components. I, A, and E represent the US images, IMU acceleration, and IMU Euler angle in the TMS's input, respectively. The best results are shown in blue.}
\centering
\begin{tabular}{p{25pt}|p{78pt}|p{50pt}|p{50pt}|p{50pt}|p{70pt}|p{50pt}|p{45pt}}
\hline\hline
\multicolumn{1}{c|}{Dataset}& \multicolumn{1}{|c|}{Method}& \multicolumn{1}{|c|}{FDR(\%)$\downarrow$}& \multicolumn{1}{|c|}{ADR(\%)$\downarrow$}& \multicolumn{1}{|c|}{MD(mm)$\downarrow$}& \multicolumn{1}{|c|}{SD(mm)$\downarrow$}& \multicolumn{1}{|c|}{HD(mm)$\downarrow$}& \multicolumn{1}{|c}{MEA(deg)$\downarrow$} \\
\hline
\multirow{9}*{Arm}& TMS/I& $18.30(12.29)$& $27.07(17.22)$& $46.87(32.33)$& $1586.31(1932.69)$& $44.13(31.09)$& $6.28(4.30)$ \\
~& TMS/IA& $17.26(9.56)$& $26.08(14.01)$& $45.72(26.29)$& $1542.28(1854.57)$& $42.40(25.73)$& $6.01(3.79)$ \\
~& TMS/IE& $16.97(9.57)$& $24.85(10.27)$& $40.30(24.73)$& $1358.24(1852.21)$& $35.99(27.16)$& $4.67(2.77)$ \\
~& TMS/IAE& \textcolor{blue}{$\mathbf{16.27(10.27)}$}& \textcolor{blue}{$\mathbf{23.23(10.26)}$}& \textcolor{blue}{$\mathbf{36.68(22.20)}$}& \textcolor{blue}{$\mathbf{1149.10(1001.49)}$}& \textcolor{blue}{$\mathbf{33.04(18.58)}$}& \textcolor{blue}{$\mathbf{4.33(2.43)}$} \\
\cline{2-8}
~& SVC& $13.32(6.81)$& $19.97(14.00)$& $31.35(17.72)$& $988.29(766.81)$& $28.21(13.96)$& $4.23(2.56)$ \\
~& PAC& $15.63(9.81)$& $22.50(12.22)$& $34.07(19.32)$& $1108.53(984.39)$& $31.69(15.82)$& $4.25(2.47)$ \\
~& PMC& $15.04(9.44)$& $21.87(11.40)$& $33.70(19.69)$& $1077.88(946.79)$& $30.33(15.17)$& $3.91(2.46)$ \\
~& MSS (MoNet)& $14.38(8.67)$& $21.20(10.44)$& $32.36(18.71)$& $1009.60(863.52)$& $28.96(14.17)$& $3.70(2.30)$ \\
~& MoNetV2& \textcolor{blue}{$\mathbf{11.04(6.02)}$}& \textcolor{blue}{$\mathbf{18.26(9.10)}$}& \textcolor{blue}{$\mathbf{28.84(15.62)}$}& \textcolor{blue}{$\mathbf{901.87(658.43)}$}& \textcolor{blue}{$\mathbf{26.24(12.47)}$}& \textcolor{blue}{$\mathbf{3.00(2.25)}$} \\
\hline
\multirow{9}*{Carotid}& TMS/I& $19.88(17.40)$& $32.21(21.34)$& $32.20(14.94)$& $898.44(606.06)$& $27.20(11.90)$& $4.55(2.20)$ \\
~& TMS/IA& $17.91(11.66)$& $29.98(16.69)$& $29.50(14.49)$& $862.01(558.00)$& $25.62(12.14)$& $4.44(2.59)$ \\
~& TMS/IE& $17.04(10.86)$& $28.37(17.30)$& $28.45(13.03)$& $861.62(828.46)$& $25.24(12.33)$& $3.87(2.13)$ \\
~& TMS/IAE& \textcolor{blue}{$\mathbf{16.37(10.98)}$}& \textcolor{blue}{$\mathbf{27.25(19.25)}$}& \textcolor{blue}{$\mathbf{25.56(11.41)}$}& \textcolor{blue}{$\mathbf{817.70(738.46)}$}& \textcolor{blue}{$\mathbf{22.97(11.44)}$}& \textcolor{blue}{$\mathbf{3.58(1.76)}$} \\
\cline{2-8}
~& SVC& $14.26(8.95)$& $25.98(18.75)$& $23.33(10.66)$& $756.36(654.01)$& $20.76(10.77)$& $3.46(1.97)$ \\
~& PAC& $15.69(10.92)$& $26.73(19.79)$& $24.85(11.04)$& $781.76(696.32)$& $22.49(11.11)$& $3.44(1.85)$ \\
~& PMC& $15.28(10.08)$& $26.48(19.08)$& $24.60(11.21)$& $792.69(707.81)$& $22.04(11.27)$& $3.05(1.86)$ \\
~& MSS (MoNet)& $14.53(9.51)$& $26.50(19.23)$& $23.67(10.69)$& $753.40(593.38)$& $21.11(10.95)$& $2.92(1.76)$ \\
~& MoNetV2& \textcolor{blue}{$\mathbf{12.11(7.55)}$}& \textcolor{blue}{$\mathbf{23.78(17.51)}$}& \textcolor{blue}{$\mathbf{20.75(8.82)}$}& \textcolor{blue}{$\mathbf{646.00(455.03)}$}& \textcolor{blue}{$\mathbf{18.50(8.64)}$}& \textcolor{blue}{$\mathbf{2.40(1.55)}$} \\
\hline
\multirow{9}*{Thyroid}& TMS/I& $24.01(13.99)$& $36.84(18.68)$& $12.01(9.46)$& $327.58(369.98)$& $10.49(7.55)$& $2.12(1.25)$ \\
~& TMS/IA& $23.28(14.10)$& $34.56(16.60)$& $11.37(9.08)$& $301.47(355.80)$& $10.41(6.47)$& $2.06(1.45)$ \\
~& TMS/IE& $22.80(13.26)$& $33.53(10.65)$& $11.05(7.41)$& $270.57(276.42)$& $10.31(5.51)$& $1.83(1.01)$ \\
~& TMS/IAE& \textcolor{blue}{$\mathbf{22.09(11.24)}$}& \textcolor{blue}{$\mathbf{32.52(13.88)}$}& \textcolor{blue}{$\mathbf{10.66(6.30)}$}& \textcolor{blue}{$\mathbf{233.61(215.06)}$}& \textcolor{blue}{$\mathbf{10.15(6.29)}$}& \textcolor{blue}{$\mathbf{1.79(1.17)}$} \\
\cline{2-8}
~& SVC& $19.18(9.17)$& $29.14(11.02)$& $9.72(4.85)$& $223.48(206.82)$& $9.24(4.78)$& $1.75(1.16)$ \\
~& PAC& $21.21(11.41)$& $31.21(11.38)$& $10.44(4.98)$& $229.66(225.69)$& $9.72(4.93)$& $1.73(1.17)$ \\
~& PMC& $20.37(10.13)$& $31.26(12.35)$& $10.05(5.52)$& $228.59(224.37)$& $9.50(5.49)$& $1.60(1.17)$ \\
~& MSS (MoNet)& $19.64(9.38)$& $29.17(11.36)$& $10.16(5.87)$& $228.12(210.99)$& $9.63(5.82)$& $1.56(1.16)$ \\
~& MoNetV2& \textcolor{blue}{$\mathbf{17.88(9.10)}$}& \textcolor{blue}{$\mathbf{26.34(9.12)}$}& \textcolor{blue}{$\mathbf{8.93(4.79)}$}& \textcolor{blue}{$\mathbf{207.60(195.20)}$}& \textcolor{blue}{$\mathbf{8.43(4.73)}$}& \textcolor{blue}{$\mathbf{1.36(0.97)}$} \\
\hline\hline
\end{tabular}
\label{tab2}
\end{table*}

Table~\ref{tab1} summarizes the overall comparison of the MoNetV2 with CNN-OF, ResNet18, DC$^2$-Net, RecON, MoNet, OSCNet, and TMS/I (TMS with images as input) on arm, carotid, and thyroid datasets. 
It should be noted that CNN-OF, which integrates IMU Euler angle, exhibits superior performance in the MEA metric compared to other sensorless methods (ResNet18, DC$^2$-Net, RecON, and TMS/I), validating the efficacy of IMU integration. However, due to its simplistic architecture, CNN-OF demonstrates the lowest performance across other metrics.
DC$^2$-Net outperforms both CNN-OF and ResNet18 in all metrics except MEA, even surpassing the temporal-based TMS/I in FDR and ADR for arm and carotid datasets. DC$^2$-Net's self-attention and contrastive learning can efficiently extract image motion information across diverse scanning velocities or tactics. 
We note that online-learning-based methods (RecON, MoNet, OSCNet, and MoNetV2) significantly outperformed others (CNN-OF, ResNet18, DC$^2$-Net, and TMS/I) based on routine training and test strategies. This suggests that online learning can effectively adapt to unseen data. 
MoNet and OSCNet surpass RecON by fully integrating IMU acceleration and Euler angle.
Undoubtedly, the MoNetV2 significantly outperformed all other methods across all datasets and metrics ($p<0.05$, $t$-test). This benefits from the velocity-based image and motion integration (TMS) that enhance transformation estimation. The MCC and MSS also adapt MoNetV2 to varying scanning velocities and tactics.

\begin{table*}[t]
\caption{Mean(std) results of the generalizability performance of different methods. The best results are shown in blue.}
\centering
\begin{tabular}{p{65pt}|p{40pt}|p{50pt}|p{50pt}|p{53pt}|p{68pt}|p{53pt}|p{45pt}}
\hline\hline
\multicolumn{1}{c|}{Train/Test set}& \multicolumn{1}{|c|}{Method}& \multicolumn{1}{|c|}{FDR(\%)$\downarrow$}& \multicolumn{1}{|c|}{ADR(\%)$\downarrow$}& \multicolumn{1}{|c|}{MD(mm)$\downarrow$}& \multicolumn{1}{|c|}{SD(mm)$\downarrow$}& \multicolumn{1}{|c|}{HD(mm)$\downarrow$}& \multicolumn{1}{|c}{MEA(deg)$\downarrow$} \\
\hline
\multirow{7}*{Arm/Carotid}& CNN-OF& $47.43(24.95)$& $63.10(44.37)$& $64.99(32.15)$& $2145.66(1610.83)$& $52.47(26.17)$& $4.38(1.09)$ \\
~& ResNet18& $52.37(34.72)$& $72.01(43.32)$& $85.54(39.76)$& $2659.89(2236.81)$& $61.98(26.89)$& $6.85(2.99)$ \\
~& DC$^2$-Net& $52.84(34.29)$& $68.40(30.42)$& $84.01(33.16)$& $2548.94(1856.93)$& $64.08(24.95)$& $7.05(3.42)$ \\
~& RecON& $35.85(26.64)$& $45.25(24.14)$& $59.64(41.83)$& $1703.25(1387.28)$& $40.33(26.75)$& $4.55(2.41)$ \\
~& MoNet& $33.65(32.96)$& $46.51(29.21)$& $54.36(28.13)$& $1589.73(1190.25)$& $39.56(24.85)$& $3.85(2.23)$ \\
~& OSCNet& $32.84(30.67)$& $46.26(30.54)$& $51.24(37.45)$& $1442.39(1145.05)$& $38.24(24.93)$& $3.62(2.07)$ \\
~& MoNetV2& \textcolor{blue}{$\mathbf{22.93(17.67)}$}& \textcolor{blue}{$\mathbf{37.40(20.91)}$}& \textcolor{blue}{$\mathbf{38.27(24.12)}$}& \textcolor{blue}{$\mathbf{1089.95(790.26)}$}& \textcolor{blue}{$\mathbf{30.95(20.11)}$}& \textcolor{blue}{$\mathbf{3.41(1.67)}$} \\
\hline
\multirow{7}*{Arm/Thyroid}& CNN-OF& $44.27(26.76)$& $59.50(37.67)$& $20.86(11.79)$& $442.58(419.24)$& $15.80(8.44)$& $2.94(1.74)$ \\
~& ResNet18& $82.43(47.57)$& $105.00(50.19)$& $38.53(24.04)$& $912.17(1074.68)$& $27.01(14.86)$& $4.15(1.89)$ \\
~& DC$^2$-Net& $63.43(35.15)$& $76.01(42.12)$& $29.54(16.35)$& $650.65(639.35)$& $20.82(10.86)$& $3.82(2.43)$ \\
~& RecON& $38.10(25.77)$& $46.22(35.76)$& $17.27(12.72)$& $413.05(550.33)$& \textcolor{blue}{$\mathbf{12.04(6.99)}$}& $3.59(2.00)$ \\
~& MoNet& $35.21(23.66)$& $45.92(30.65)$& $15.68(8.05)$& $415.23(505.96)$& $13.85(7.32)$& $2.82(1.60)$ \\
~& OSCNet& $33.41(22.39)$& $48.67(28.59)$& $15.11(7.64)$& $419.16(494.05)$& $13.23(7.28)$& $2.73(1.50)$ \\
~& MoNetV2& \textcolor{blue}{$\mathbf{30.78(20.13)}$}& \textcolor{blue}{$\mathbf{41.14(22.78)}$}& \textcolor{blue}{$\mathbf{14.15(8.21)}$}& \textcolor{blue}{$\mathbf{324.22(320.53)}$}& $13.00(7.90)$& \textcolor{blue}{$\mathbf{2.40(1.25)}$} \\
\hline
\multirow{7}*{Carotid/Arm}& CNN-OF& $47.02(20.09)$& $51.22(19.08)$& $115.97(39.13)$& $3393.49(2119.88)$& $108.73(37.21)$& $5.64(4.27)$ \\
~& ResNet18& $64.06(30.41)$& $70.90(27.53)$& $162.22(66.27)$& $4833.68(3329.25)$& $149.38(58.83)$& $8.41(4.05)$ \\
~& DC$^2$-Net& $57.23(27.41)$& $63.50(24.96)$& $145.29(60.26)$& $4342.74(2999.68)$& $133.62(52.71)$& $8.63(4.16)$ \\
~& RecON& $37.38(19.10)$& $41.26(17.92)$& $93.86(41.00)$& $2744.06(1909.03)$& $82.85(32.88)$& $6.01(3.92)$ \\
~& MoNet& $32.41(19.26)$& $36.91(19.83)$& $75.96(34.26)$& $2356.11(1392.15)$& $69.37(27.13)$& $5.02(3.81)$ \\
~& OSCNet& $30.44(19.46)$& $33.67(19.32)$& $71.22(29.77)$& $2043.35(1305.22)$& $64.30(25.04)$& $4.87(3.77)$ \\
~& MoNetV2& \textcolor{blue}{$\mathbf{26.96(14.73)}$}& \textcolor{blue}{$\mathbf{29.29(14.89)}$}& \textcolor{blue}{$\mathbf{64.32(22.70)}$}& \textcolor{blue}{$\mathbf{1831.83(1064.41)}$}& \textcolor{blue}{$\mathbf{61.34(21.39)}$}& \textcolor{blue}{$\mathbf{4.49(2.81)}$} \\
\hline
\multirow{7}*{Carotid/Thyroid}& CNN-OF& $52.90(25.02)$& $63.59(27.47)$& $23.83(10.07)$& $532.06(496.67)$& $21.42(10.01)$& $2.30(1.27)$ \\
~& ResNet18& $73.82(33.02)$& $85.64(35.38)$& $34.01(16.49)$& $770.13(709.83)$& $27.50(13.67)$& $4.68(1.43)$ \\
~& DC$^2$-Net& $67.90(33.31)$& $86.67(36.62)$& $31.45(16.39)$& $730.83(654.63)$& $25.42(13.46)$& $4.53(1.40)$ \\
~& RecON& $42.70(21.40)$& $47.76(21.49)$& $19.54(10.70)$& $448.10(424.73)$& $15.25(7.78)$& $2.97(1.29)$ \\
~& MoNet& $39.56(31.02)$& $48.21(32.29)$& $16.51(8.63)$& $403.52(443.10)$& $13.95(7.19)$& $2.26(1.26)$ \\
~& OSCNet& $38.95(32.47)$& $48.86(34.01)$& $15.08(8.01)$& $373.12(451.76)$& $13.55(6.89)$& $2.22(1.16)$ \\
~& MoNetV2& \textcolor{blue}{$\mathbf{31.34(19.74)}$}& \textcolor{blue}{$\mathbf{46.53(31.04)}$}& \textcolor{blue}{$\mathbf{13.58(5.98)}$}& \textcolor{blue}{$\mathbf{327.40(336.07)}$}& \textcolor{blue}{$\mathbf{12.50(6.02)}$}& \textcolor{blue}{$\mathbf{2.08(1.40)}$} \\
\hline
\multirow{7}*{Thyroid/Arm}& CNN-OF& $48.26(20.04)$& $51.06(17.16)$& $116.89(35.42)$& $3417.76(1938.59)$& $111.30(36.50)$& $5.07(2.65)$ \\
~& ResNet18& $67.29(28.57)$& $72.55(23.64)$& $166.22(54.05)$& $4846.17(2911.39)$& $161.56(53.27)$& $8.36(3.83)$ \\
~& DC$^2$-Net& $60.60(25.20)$& $68.00(22.07)$& $160.31(51.68)$& $4533.54(2717.94)$& $151.81(50.72)$& $8.32(3.83)$ \\
~& RecON& $37.97(16.80)$& $41.05(14.58)$& $94.01(31.88)$& $2727.79(1606.68)$& $86.37(29.03)$& $5.88(3.96)$ \\
~& MoNet& $34.81(15.62)$& $36.22(13.72)$& $85.43(30.11)$& $2581.71(1592.37)$& $80.12(27.69)$& $5.34(2.00)$ \\
~& OSCNet& $33.32(14.99)$& $34.75(13.43)$& $82.20(29.71)$& $2394.12(1521.16)$& $77.63(27.04)$& $5.28(1.94)$ \\
~& MoNetV2& \textcolor{blue}{$\mathbf{30.77(13.52)}$}& \textcolor{blue}{$\mathbf{32.83(12.56)}$}& \textcolor{blue}{$\mathbf{76.87(27.82)}$}& \textcolor{blue}{$\mathbf{2229.63(1352.38)}$}& \textcolor{blue}{$\mathbf{74.14(26.36)}$}& \textcolor{blue}{$\mathbf{4.89(2.88)}$} \\
\hline
\multirow{7}*{Thyroid/Carotid}& CNN-OF& $39.23(20.97)$& $48.45(23.06)$& $61.07(19.53)$& $1878.51(1238.16)$& $56.12(19.80)$& $4.64(3.86)$ \\
~& ResNet18& $50.79(22.88)$& $66.41(23.80)$& $78.70(23.50)$& $2612.51(2204.20)$& $63.02(24.88)$& $6.13(2.77)$ \\
~& DC$^2$-Net& $48.14(17.71)$& $59.83(19.11)$& $74.96(21.14)$& $2558.02(1760.07)$& $52.79(20.84)$& $6.82(3.62)$ \\
~& RecON& $31.73(20.25)$& $41.74(21.68)$& $50.24(21.19)$& \textcolor{blue}{$\mathbf{1378.75(710.87)}$}& $43.23(17.45)$& $4.83(2.02)$ \\
~& MoNet& $30.91(18.02)$& $39.23(15.94)$& $50.62(24.36)$& $1557.29(1134.26)$& $46.23(19.16)$& $3.81(3.01)$ \\
~& OSCNet& $30.16(18.11)$& $38.65(15.65)$& $50.53(24.04)$& $1574.42(1198.33)$& $45.33(18.87)$& $3.58(2.72)$ \\
~& MoNetV2& \textcolor{blue}{$\mathbf{27.23(16.55)}$}& \textcolor{blue}{$\mathbf{35.99(15.10)}$}& \textcolor{blue}{$\mathbf{47.33(23.14)}$}& $1411.34(987.16)$& \textcolor{blue}{$\mathbf{42.25(19.72)}$}& \textcolor{blue}{$\mathbf{3.28(2.86)}$} \\
\hline\hline
\end{tabular}
\label{tabgen}
\end{table*}

Fig.~\ref{fig9} illustrates typical reconstruction cases. The MoNetV2 produces results that are closest to the ground truth. Conversely, other methods show notable discrepancies from the ground truth, such as compressed or elongated length, lateral drift, and inaccurate directional estimation. MoNetV2 has the lowest reconstruction error and the most accurate trajectory estimation. It accurately captures anatomical details such as the structure of blood vessels.
Fig.~\ref{fig10} shows the declining metric curves of all metrics for online iterative optimization. As the iterations increase, almost all metrics decrease and tend to converge, indicating that MCC and MSS can achieve effective improvement by leveraging latent cues present in the test scans. Compared to RecON, MoNet, and OSCNet, MoNetV2 achieved the lowest final metrics. The FDR and MEA metric exhibit the most prominent reduction, decreasing by $32.14\%$/$26.02\%$/$19.05\%$ and $30.71\%$/$32.96\%$/$24.02\%$ on the arm/carotid/thyroid scans, respectively.

\subsection{Ablation Studies on Inputs and Components}
\label{sec:results:ablation}
We conducted an ablation study of MoNetV2 to ascertain the effectiveness of various inputs and components, with comprehensive results detailed in Table~\ref{tab2}. In terms of inputs, we analyze several input configurations of our TMS, including US images alone (I), US images with IMU acceleration (IA), US images with IMU Euler angle (IE), and US images with both IMU acceleration and Euler angle (IAE). 
Each input variant produces a corresponding adaptation within the motion branch of the TMS. For instance, input I removes the entire motion branch, and input IE excludes the branch for acceleration within the motion branch. 
As expected, the addition of either acceleration (IA) or Euler angle (IE) yields a notable enhancement in performance over US images alone (I). 
Interestingly, even though the acceleration (IA) contributes positively, it falls short of the improvements realized through the Euler angle (IE). This outcome reinforces the notion that while low-SNR acceleration may exhibit noise at discrete points, it reliably maintains a correct trend across a broader scope.
Integrating both acceleration and Euler angle (IAE) enhances the performance further. Compared to input I, input IAE reduces FDR/ADR/MEA metrics by $11.09\%$/$14.19\%$/$31.05\%$, $17.66\%$/$15.40\%$/$21.32\%$, and $8.00\%$/$11.73\%$/$15.57\%$ across the arm, carotid, and thyroid datasets, respectively.

In the ablation study on various components, we focus on the SVC, PAC, PMC, and MSS. As shown in Table~\ref{tab2}, incorporating each individual component leads to enhanced reconstruction performance across all datasets. Compared to TMS with the US image and IMU data (TMS/IAE), the average percentage decreases in FDR/ADR/MEA metrics across the three datasets are $14.73\%$/$9.70\%$/$2.63\%$ for SVC, $4.02\%$/$3.03\%$/$3.04\%$ for PAC, $7.33\%$/$4.18\%$/$11.70\%$ for PMC, and $11.31\%$/$7.26\%$/$15.27\%$ for MSS. The improvements achieved through SVC, PAC, and PMC confirm the effectiveness of leveraging multi-level consistency to guide inter-frame transformation estimation. Meanwhile, the improvement of MSS underscores the advantages of exploiting multi-modal motion information.
Our proposed MoNetV2 significantly enhances reconstruction performance compared to adding each individual component ($p<0.05$, $t$-test). This demonstrates the effectiveness of combining SVC, PAC, PMC, and MSS.

\begin{figure}[t]
\centerline{\includegraphics[width=\columnwidth]{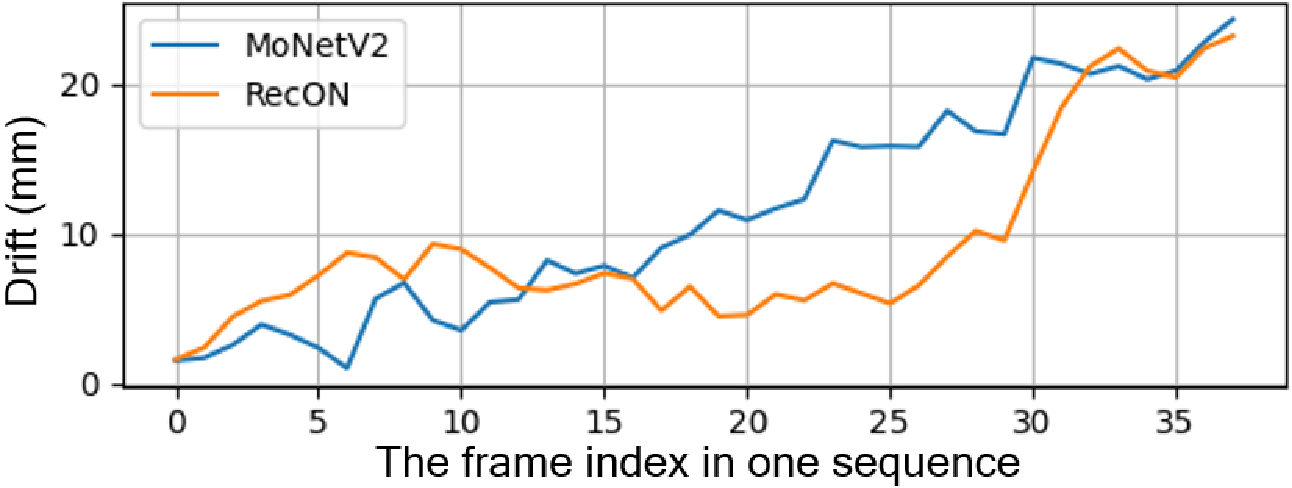}}
\caption{Typical drift curves for MoNetV2 and RecON in the thyroid/carotid setting in generalizability study.}
\label{fig:sd}
\end{figure} 

\begin{figure*}[t]
\centerline{\includegraphics[width=0.93\textwidth]{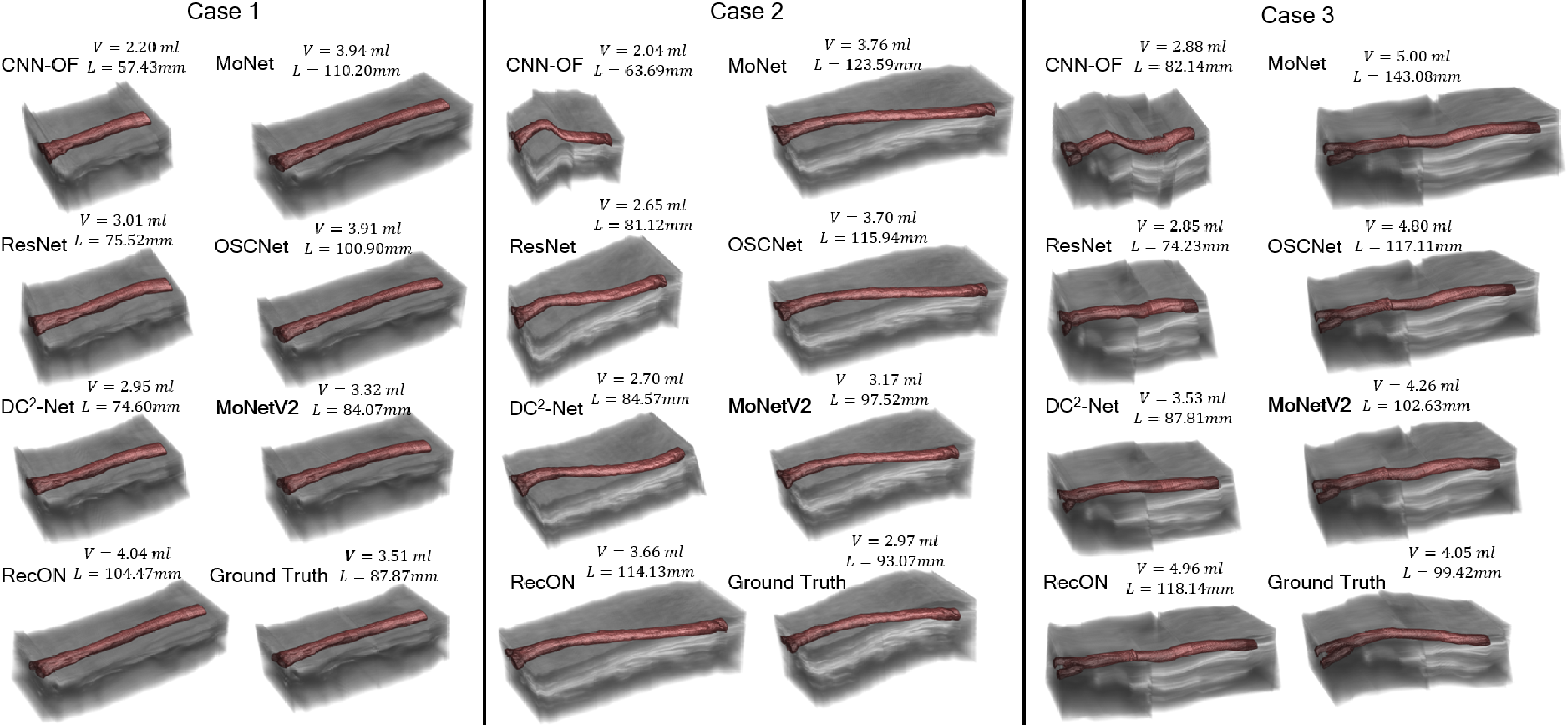}}
\caption{Three cases of vessel segmentation (brown) from the real or reconstructed volume (gray). $V$ and $L$ represent the vessel volume size and length, respectively.}
\label{fig13}
\end{figure*}

\begin{figure*}[t]
\centerline{\includegraphics[width=0.93\textwidth]{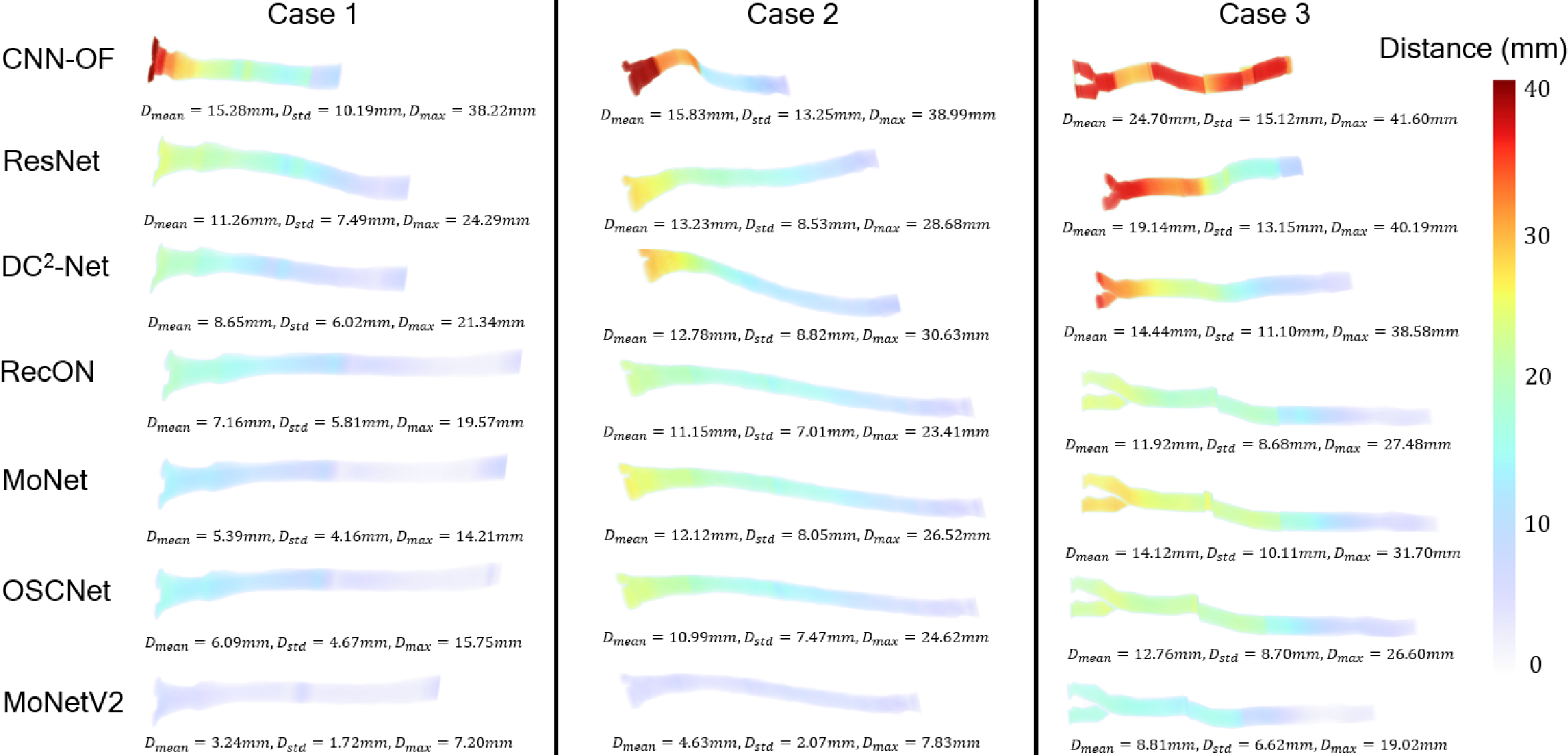}}
\caption{Voxel-to-voxel distance models between vessel segmentation from the real and reconstructed volume. Lower distances yield better performance. $D_{mean}$, $D_{std}$, and $D_{max}$ represent the mean, standard deviation of, and maximum voxel-to-voxel distance, respectively.}
\label{fig13_1}
\end{figure*}

\subsection{Generalizability Study}
\label{sec:results:generalizability}
We analyze the generalizability of all methods across the three datasets. Generalizability refers to the ability of a method on unseen data with different distributions~\cite{song2022learning}. A superior method should have better generalizability performance. We trained each method on one dataset and subsequently evaluated its performance on the others. The results of the generalizability study are summarized in Table~\ref{tabgen}. The performance of all methods decreased compared to the results shown in Table~\ref{tab1}, but MoNetV2 still surpasses the other methods on almost all metrics. This is attributed to MoNetV2's motion integration (TMS), multi-level consistency (SVC, PAC, and PMC), and multi-modal strategy (MSS). In addition, online-learning-based methods (RecON, MoNet, and OSCNet) continue to show superior performance compared to methods using routine training and test strategies (CNN-OF, ResNet18, and DC$^2$-Net), confirming the advantages of online learning for handling unseen data. In contrast to Table~\ref{tab1}, CNN-OF shows superior generalizability to ResNet18 and DC$^2$-Net. This phenomenon also appears in Luo et al.~\cite{luo2023recon}, which is attributed to CNN-OF relying on the common speckle pattern.

An interesting phenomenon in Table~\ref{tabgen} is that RecON obtained slightly lower SD values than MoNetV2 in the thyroid/carotid setting. During the thyroid and carotid US scanning, the tissue structure is easily deformed due to compression, resulting in significant scanning jitter. RecON considers complex scanning techniques but does not incorporate IMU data. Compared to MoNetV2, RecON exhibits greater fluctuations in the frame-to-frame distance error (estimated distance minus actual distance) across the entire scanning sequence under conditions of significant jitter. The frame-to-frame distance error at different frames in the sequence partially cancels out~\cite{luo2023recon}, making the SD of RecON relatively smaller while other metrics are relatively larger. Fig.~\ref{fig:sd} provides a specific example of lower SD phenomenon, where the area under the drift curve represents the SD value. It can be seen that RecON's drift is declining in the middle of the curve due to error cancellation. Since jitter is particularly pronounced in thyroid scanning compared to carotid scanning, the lower SD situation occurs with a relatively high probability in the thyroid/carotid setting. Therefore, RecON’s mean SD is slightly lower than that of MoNetV2.

\subsection{Reconstruction of Blood Vessels}
\label{sec:results:vessel}

\begin{figure}[t]
\begin{subfigure}{\linewidth}
  \centering
  \includegraphics[width=0.94\linewidth]{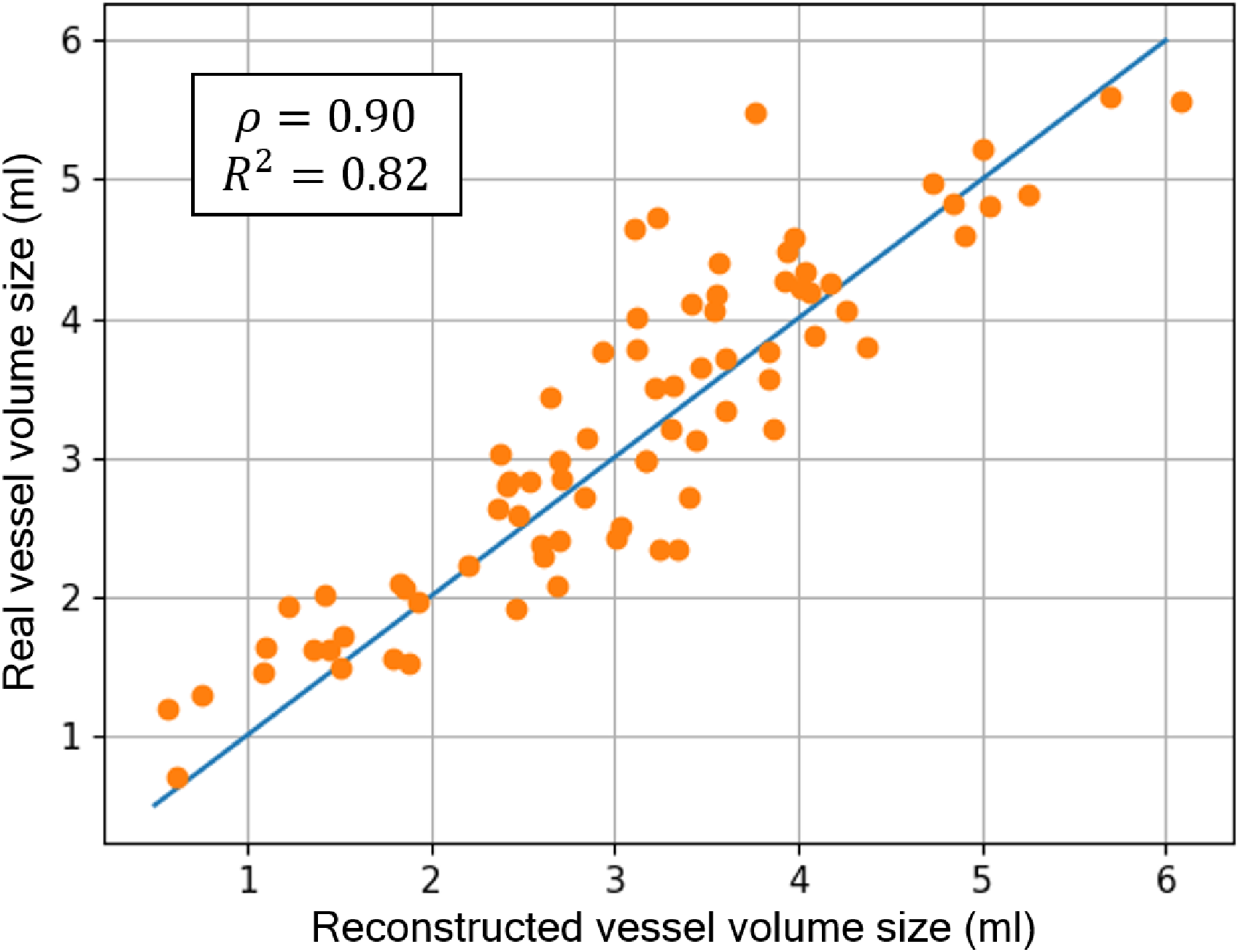}  
  \caption{Correlation of the reconstructed and real vessel volume sizes.}
  \label{fig_15_1}
\end{subfigure}
\begin{subfigure}{\linewidth}
  \centering
  \includegraphics[width=0.97\linewidth]{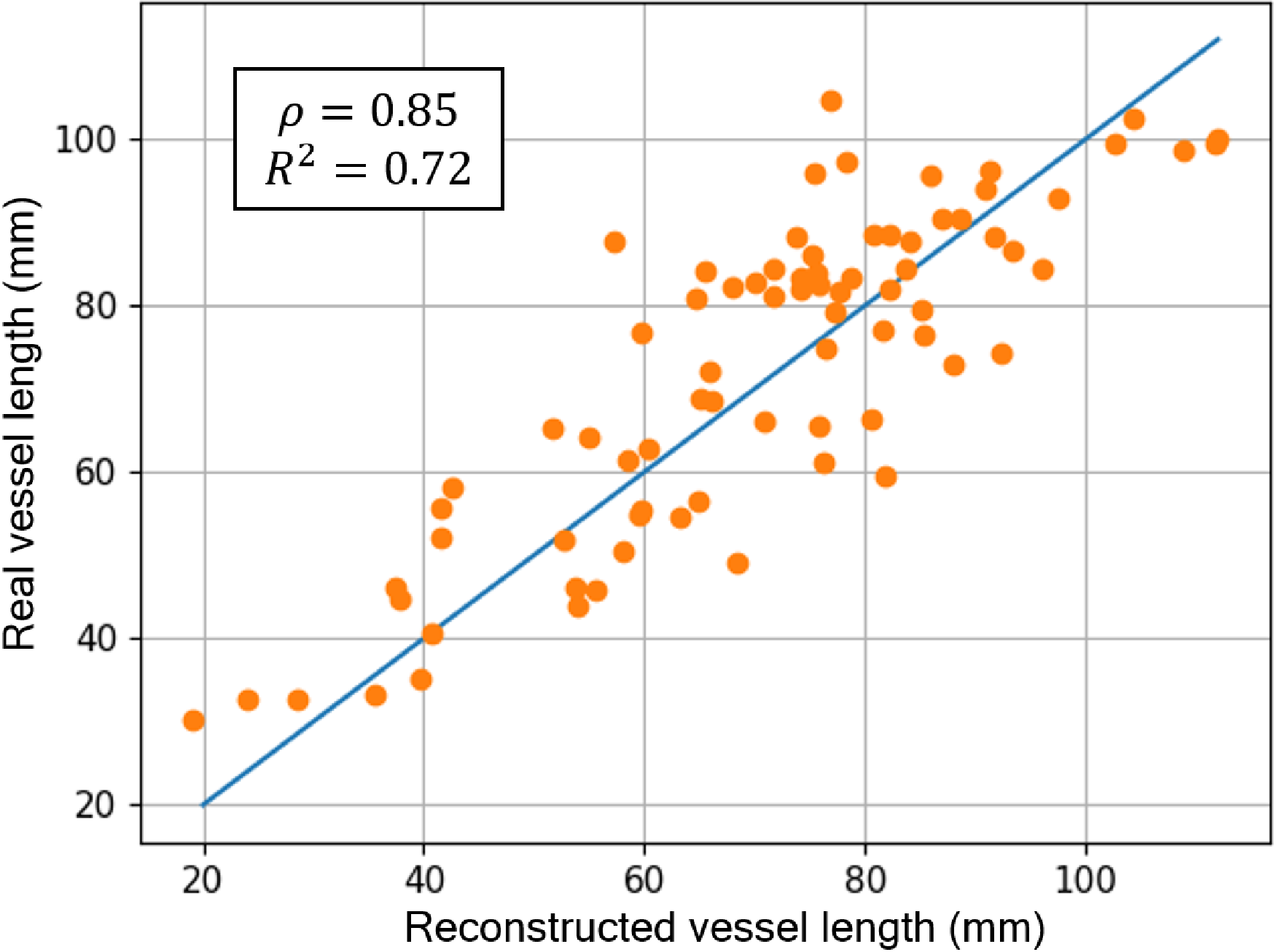}  
  \caption{Correlation of the reconstructed and real vessel lengths.}
  \label{fig_15_2}
\end{subfigure}
\caption{Correlation of the vessel volume sizes (a) and lengths (b) between the reconstructed and real carotid volumes. Each orange dot indicates a sample of the vessel volume size (a) or length (b). The blue line represents the diagonal line.}
\label{fig15}
\end{figure}

\begin{figure}[t]
\begin{subfigure}{\linewidth}
  \centering
  \includegraphics[width=1.0\linewidth]{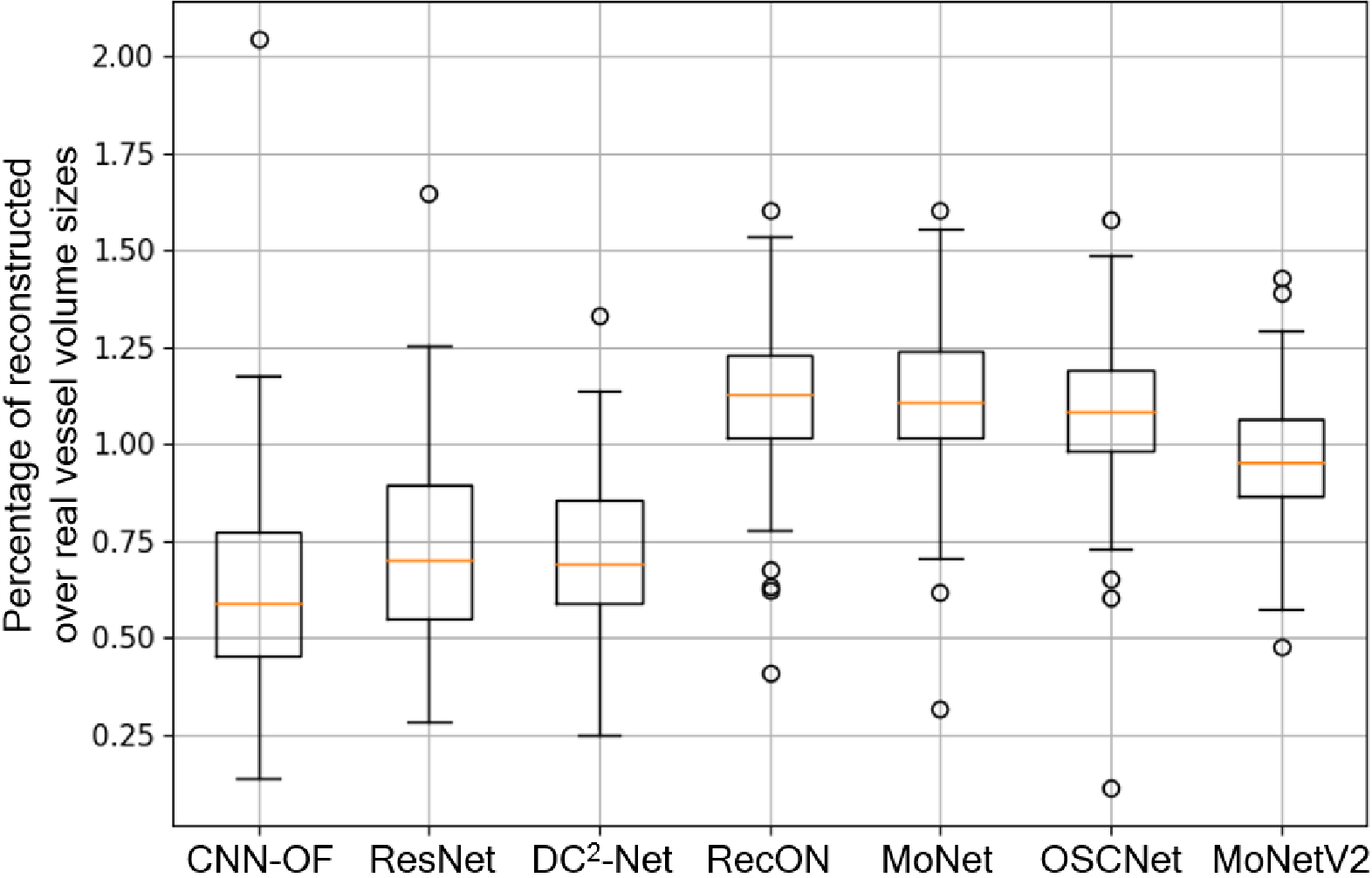}  
  \caption{Boxplot of the percentage of reconstructed over real vessel volume sizes.}
  \label{fig_16_1}
\end{subfigure}
\begin{subfigure}{\linewidth}
  \centering
  \includegraphics[width=1.0\linewidth]{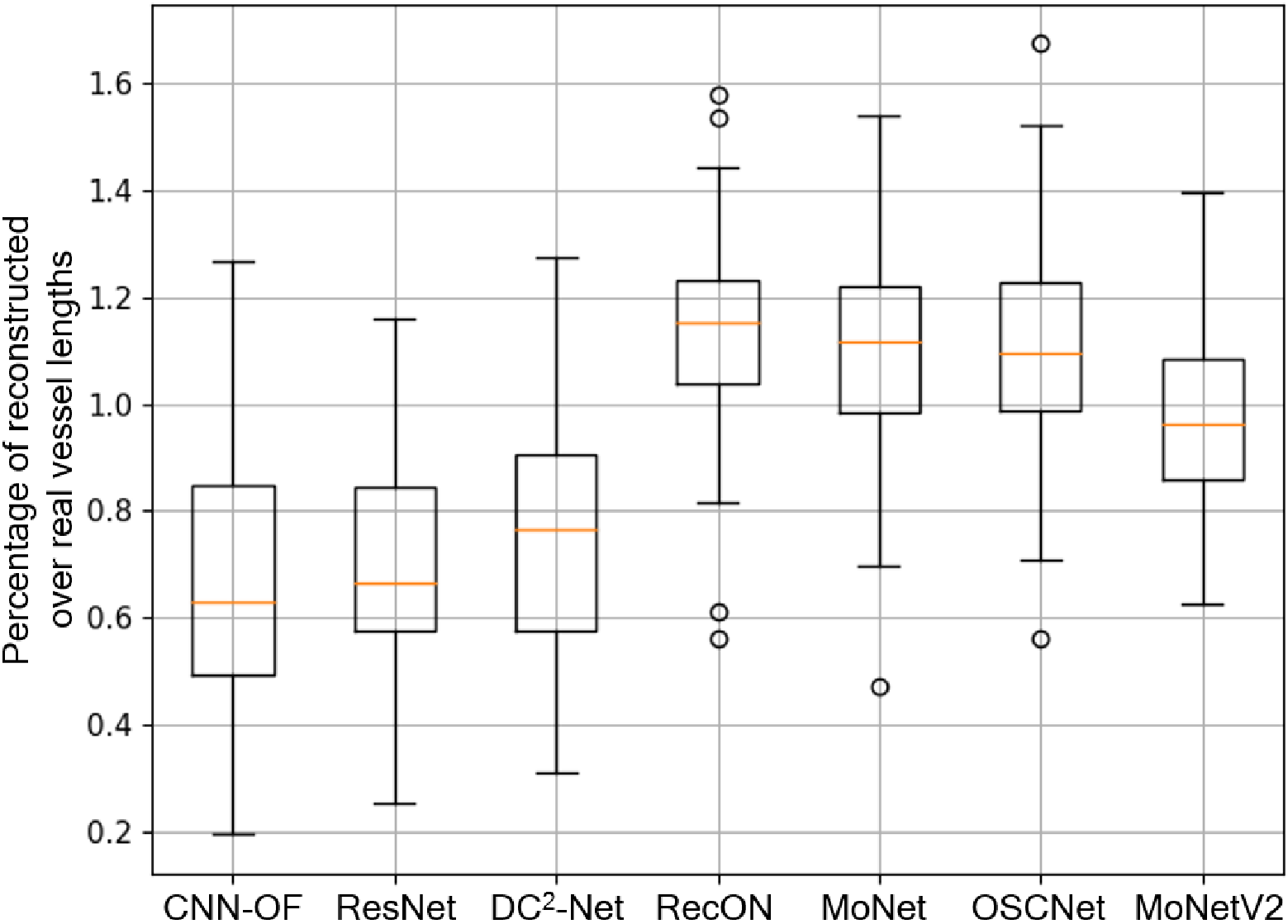}  
  \caption{Boxplot of the percentage of reconstructed over real vessel lengths.}
  \label{fig_16_2}
\end{subfigure}
\caption{Boxplot of the percentage of reconstructed over real vessel volume sizes (a) and lengths (b) on different methods. The boxplot includes the minimum, first quartile (Q1), median, third quartile (Q3), and maximum values. The x- and y-axis represent the methods and percentages, respectively.}
\label{fig16}
\end{figure}

Vessel segmentation quantifies the morphological features of blood vessels, such as their diameter, length, and branching patterns. It is crucial for vessel pathology analysis, disease diagnosis, and treatment progression.
We conduct vessel segmentation on the carotid volume to further evaluate MoNetV2. Specifically, for the 78 test scans in the carotid dataset, we utilize the estimated and actual inter-frame transformation parameters to obtain reconstructed and real carotid volumes, respectively. Subsequently, we use the Pair software~\cite{liang2022sketch} to semi-automatically label vessel masks in the carotid volumes. 
The length of blood vessels is calculated as the sequential connection of center points in the masks on all frames in 3D space.
Excellent reconstruction performance ensures consistency between vessel segmentation inside the reconstructed and real volumes. Therefore, we qualitatively and quantitatively assess MoNetV2's reconstruction performance by the morphology, volume size, and length of the vessels.

\begin{table*}[t]
\caption{Mean(std) results of the reconstruction performance of different architectures in TMS. The best results are shown in blue.}
\centering
\begin{tabular}{p{25pt}|p{67pt}|p{50pt}|p{50pt}|p{50pt}|p{70pt}|p{50pt}|p{45pt}}
\hline\hline
\multicolumn{1}{c|}{Dataset}& \multicolumn{1}{|c|}{Architecture}& \multicolumn{1}{|c|}{FDR(\%)$\downarrow$}& \multicolumn{1}{|c|}{ADR(\%)$\downarrow$}& \multicolumn{1}{|c|}{MD(mm)$\downarrow$}& \multicolumn{1}{|c|}{SD(mm)$\downarrow$}& \multicolumn{1}{|c|}{HD(mm)$\downarrow$}& \multicolumn{1}{|c}{MEA(deg)$\downarrow$} \\
\hline
\multirow{3}*{Arm}& ResNet18& \textcolor{blue}{$\mathbf{16.27(10.27)}$}& \textcolor{blue}{$\mathbf{23.23(10.26)}$}& \textcolor{blue}{$\mathbf{36.68(22.20)}$}& \textcolor{blue}{$\mathbf{1149.10(1001.49)}$}& \textcolor{blue}{$\mathbf{33.04(18.58)}$}& \textcolor{blue}{$\mathbf{4.33(2.43)}$} \\
~& ResNeXt50& $21.49(13.19)$& $32.41(19.76)$& $53.97(26.65)$& $1580.98(1689.22)$& $47.83(28.83)$& $5.33(3.25)$ \\
~& MobileNetV4& $18.46(12.37)$& $26.41(17.01)$& $39.65(24.86)$& $1278.53(1356.91)$& $38.23(25.34)$& $5.18(2.96)$ \\
\hline
\multirow{3}*{Carotid}& ResNet18& \textcolor{blue}{$\mathbf{16.37(10.98)}$}& \textcolor{blue}{$\mathbf{27.25(19.25)}$}& \textcolor{blue}{$\mathbf{25.56(11.41)}$}& \textcolor{blue}{$\mathbf{817.70(738.46)}$}& \textcolor{blue}{$\mathbf{22.97(11.44)}$}& \textcolor{blue}{$\mathbf{3.58(1.76)}$} \\
~& ResNeXt50& $19.79(13.40)$& $30.83(20.41)$& $31.22(15.63)$& $905.88(763.28)$& $26.43(12.56)$& $4.62(2.67)$ \\
~& MobileNetV4& $17.89(11.86)$& $28.91(20.31)$& $27.40(13.74)$& $857.52(723.84)$& $24.58(11.65)$& $3.75(2.10)$ \\
\hline
\multirow{3}*{Thyroid}& ResNet18& \textcolor{blue}{$\mathbf{22.09(11.24)}$}& \textcolor{blue}{$\mathbf{32.52(13.88)}$}& \textcolor{blue}{$\mathbf{10.66(6.30)}$}& \textcolor{blue}{$\mathbf{233.61(215.06)}$}& \textcolor{blue}{$\mathbf{10.15(6.29)}$}& \textcolor{blue}{$\mathbf{1.79(1.17)}$} \\
~& ResNeXt50& $25.06(13.16)$& $36.85(18.91)$& $12.32(9.56)$& $295.95(313.75)$& $12.46(6.85)$& $2.21(1.35)$ \\
~& MobileNetV4& $23.14(12.38)$& $33.61(15.82)$& $10.95(7.43)$& $249.79(274.51)$& $11.46(5.92)$& $1.86(1.19)$ \\
\hline\hline
\end{tabular}
\label{tab:resnet18}
\end{table*}

\begin{table*}[t]
\caption{Mean(std) results of the reconstruction performance of different integration methods of the velocity feature in TMS. The best results are shown in blue.}
\centering
\begin{tabular}{p{25pt}|p{67pt}|p{50pt}|p{50pt}|p{50pt}|p{70pt}|p{50pt}|p{45pt}}
\hline\hline
\multicolumn{1}{c|}{Dataset}& \multicolumn{1}{|c|}{Integration Method}& \multicolumn{1}{|c|}{FDR(\%)$\downarrow$}& \multicolumn{1}{|c|}{ADR(\%)$\downarrow$}& \multicolumn{1}{|c|}{MD(mm)$\downarrow$}& \multicolumn{1}{|c|}{SD(mm)$\downarrow$}& \multicolumn{1}{|c|}{HD(mm)$\downarrow$}& \multicolumn{1}{|c}{MEA(deg)$\downarrow$} \\
\hline
\multirow{2}*{Arm}& Addition& \textcolor{blue}{$\mathbf{16.27(10.27)}$}& \textcolor{blue}{$\mathbf{23.23(10.26)}$}& \textcolor{blue}{$\mathbf{36.68(22.20)}$}& \textcolor{blue}{$\mathbf{1149.10(1001.49)}$}& \textcolor{blue}{$\mathbf{33.04(18.58)}$}& \textcolor{blue}{$\mathbf{4.33(2.43)}$} \\
~& Concatenation& $17.44(10.39)$& $24.35(10.21)$& $38.03(22.30)$& $1263.68(1046.85)$& $34.17(18.26)$& $4.34(2.43)$ \\
\hline
\multirow{2}*{Carotid}& Addition& \textcolor{blue}{$\mathbf{16.37(10.98)}$}& \textcolor{blue}{$\mathbf{27.25(19.25)}$}& \textcolor{blue}{$\mathbf{25.56(11.41)}$}& \textcolor{blue}{$\mathbf{817.70(738.46)}$}& \textcolor{blue}{$\mathbf{22.97(11.44)}$}& \textcolor{blue}{$\mathbf{3.58(1.76)}$} \\
~& Concatenation& $16.97(11.23)$& $27.52(19.34)$& $25.73(11.51)$& $835.83(746.32)$& $23.02(11.74)$& $3.68(1.78)$ \\
\hline
\multirow{2}*{Thyroid}& Addition& \textcolor{blue}{$\mathbf{22.09(11.24)}$}& \textcolor{blue}{$\mathbf{32.52(13.88)}$}& \textcolor{blue}{$\mathbf{10.66(6.30)}$}& \textcolor{blue}{$\mathbf{233.61(215.06)}$}& \textcolor{blue}{$\mathbf{10.15(6.29)}$}& \textcolor{blue}{$\mathbf{1.79(1.17)}$} \\
~& Concatenation& $24.02(12.06)$& $35.46(16.89)$& $11.99(7.65)$& $296.94(305.41)$& $11.23(6.64)$& $1.83(1.19)$ \\
\hline\hline
\end{tabular}
\label{tab:integrate}
\end{table*}

Fig.~\ref{fig13} shows the reconstructed and real carotid volumes along with their corresponding vessel segmentations. 
MoNetV2 accurately reconstructs the morphology and bifurcation, obtaining the closest vessel volume size and length to the ground truth.
In comparison, CNN-OF, ResNet18, DC$^2$-Net, RecON, MoNet, and OSCNet demonstrate relatively low reconstruction accuracy, particularly in terms of length.
We established voxel correspondence between the reconstructed and real vessels by considering the frame positions. Therefore, we calculated the voxel-to-voxel distance models for the different methods, as illustrated in Fig.~\ref{fig13_1}. MoNetV2 achieved the lowest voxel-to-voxel distances, measuring $3.24\pm 1.72$ mm, $4.63\pm 2.07$ mm, and $8.81\pm 6.62$ mm for Cases 1, 2, and 3, respectively.
In addition, we analyzed the volume size and length of the segmented vessels to evaluate the performance of MoNetV2. Fig.~\ref{fig15} illustrates the correlation of the volume sizes (subfigure~\ref{fig_15_1}) and lengths (subfigure~\ref{fig_15_2}) of vessel segmentation between the reconstructed and real carotid volumes. In each subfigure, all samples are distributed around the diagonal line. The mean errors for the volume sizes and lengths are $0.42\pm 0.33$ ml and $9.02\pm 6.26$ mm, respectively. The strong correlations ($\rho=0.90$ for vessel volume size and $\rho=0.85$ for vessel length) demonstrate that the reconstruction performance of MoNetV2 can provide valuable information for subsequent diagnostic analysis. Fig.~\ref{fig16} shows boxplots representing the percentage of reconstructed vessel volume sizes (subfigure~\ref{fig_16_1}) and vessel lengths (subfigure~\ref{fig_16_2}) over their real counterparts across various methods. A percentage closer to 1.0 indicates superior reconstruction performance. It can be observed that MoNetV2 achieves the closest result to 1.0 for both volume size and length. This demonstrates the advantages of motion information integration, multi-level consistency, and multi-modal strategy in reconstruction.

\section{Discussion and Conclusion}
\label{sec:disclu}
We propose an enhanced motion network (MoNetV2) that integrates motion information, multi-level consistency, and multi-modal strategy to address the challenges in freehand 3D US reconstruction. MoNetV2 surpasses current state-of-the-art methods consistently across a diverse range of datasets and evaluation metrics.
The introduction of the lightweight IMU sensor provides motion information beyond the US images, enhancing the accuracy and generalizability of inter-frame transformation estimation. The TMS effectively integrates the image and motion information from a velocity perspective to improve image-only reconstruction accuracy. The SVC, PAC, and PMC leverage the inherent consistency of unseen scans for adaptive optimization to handle the various scanning velocities or tactics. The MSS further reduces accumulated errors by enforcing agreement between estimated values and IMU measurements. 
MoNetV2 demonstrates superior accuracy and generalizability in complex scenarios involving different scanning velocities and tactics. 
The experimental results demonstrate the effectiveness of MoNetV2 in improving reconstruction performance compared to state-of-the-art methods. 
In the vessel segmentation experiment, MoNetV2 also shows better consistency with the ground truth. This illustrates its potential value in subsequent clinical diagnosis and analysis applications.
However, the MoNetV2 has several limitations. First, the MoNetV2 is designed for estimating rigid transformations between US frames. In practice, tissues frequently experience elastic deformations during freehand scanning due to compression or the force exerted by the probe. Intuitively, incorporating these tissue deformations into the reconstruction process enhances both realism and spatial accuracy. This necessitates modeling tissue biomechanics and integrating those models into the reconstruction pipeline. Second, the current implementation of the MoNetV2 involves online inference, with an average inference time of approximately 1.5 minutes for each scan. For the technology to be fully embraced in clinical settings, developing capabilities for real-time or near real-time processing is crucial. Achieving this level of performance will likely demand more streamlined model architectures and more efficient deployment on GPU platforms, ensuring that the system can operate seamlessly within the fast-paced environment of clinical practice.

In conclusion, we propose MoNetV2 for freehand 3D US reconstruction. By integrating sensor-based motion information, multi-level consistency, and multi-modal strategy, MoNetV2 addresses the challenges associated with relying solely on US images. It demonstrates superior accuracy and generalizability across diverse scanning velocities and tactics. In our future work, we aim to explore the integration of biomechanical modeling to simulate soft tissue deformation. Developing real-time or near real-time solutions is also a key priority going forward. We believe MoNetV2 represents a vital step toward making freehand 3D US reconstruction viable for practical clinical usage.

\begin{table*}[t]
\caption{Mean(std) results of the reconstruction performance on different numbers of interpolated images within the PMC component. The best results are shown in blue.}
\centering
\begin{tabular}{p{25pt}|p{25pt}|p{50pt}|p{50pt}|p{50pt}|p{70pt}|p{50pt}|p{45pt}}
\hline\hline
\multicolumn{1}{c|}{Dataset}& \multicolumn{1}{|c|}{Number}& \multicolumn{1}{|c|}{FDR(\%)$\downarrow$}& \multicolumn{1}{|c|}{ADR(\%)$\downarrow$}& \multicolumn{1}{|c|}{MD(mm)$\downarrow$}& \multicolumn{1}{|c|}{SD(mm)$\downarrow$}& \multicolumn{1}{|c|}{HD(mm)$\downarrow$}& \multicolumn{1}{|c}{MEA(deg)$\downarrow$} \\
\hline
\multirow{4}*{Arm}& 0& 11.36(6.05)& 18.50(9.15)& 29.25(15.72)& 915.63(670.74)& 26.44(12.59)& 3.08(2.25) \\
~& 15& 11.25(5.99)& 18.44(9.17)& 29.05(15.70)& 910.63(665.28)& 26.39(12.55)& 3.05(2.26) \\
~& 31& 11.19(5.97)& 18.36(9.13)& 28.96(15.69)& 906.92(660.18)& 26.33(12.48)& 3.03(2.24) \\
~& 63& \textcolor{blue}{$\mathbf{11.04(6.02)}$}& \textcolor{blue}{$\mathbf{18.26(9.10)}$}& \textcolor{blue}{$\mathbf{28.84(15.62)}$}& \textcolor{blue}{$\mathbf{901.87(658.43)}$}& \textcolor{blue}{$\mathbf{26.24(12.47)}$}& \textcolor{blue}{$\mathbf{3.00(2.25)}$} \\
\hline
\multirow{4}*{Carotid}& 0& 12.40(7.61)& 24.05(17.60)& 21.02(8.89)& 658.23(476.65)& 18.89(8.75)& 2.45(1.58) \\
~& 15& 12.35(7.59)& 23.95(17.54)& 20.95(8.85)& 655.19(470.23)& 18.76(8.72)& 2.44(1.57) \\
~& 31& 12.25(7.49)& 23.89(17.52)& 20.86(8.80)& 651.56(462.71)& 18.68(8.71)& 2.42(1.57) \\
~& 63& \textcolor{blue}{$\mathbf{12.11(7.55)}$}& \textcolor{blue}{$\mathbf{23.78(17.51)}$}& \textcolor{blue}{$\mathbf{20.75(8.82)}$}& \textcolor{blue}{$\mathbf{646.00(455.03)}$}& \textcolor{blue}{$\mathbf{18.50(8.64)}$}& \textcolor{blue}{$\mathbf{2.40(1.55)}$} \\
\hline
\multirow{4}*{Thyroid}& 0& 18.01(9.19)& 26.70(9.32)& 9.12(4.95)& 209.84(197.82)& 8.55(4.85)& 1.39(0.99) \\
~& 15& 17.98(9.19)& 26.60(9.21)& 9.08(4.89)& 209.12(196.76)& 8.52(4.84)& 1.38(0.99) \\
~& 31& 17.93(9.15)& 26.54(9.15)& 9.02(4.85)& 208.54(196.35)& 8.48(4.80)& 1.37(0.98) \\
~& 63& \textcolor{blue}{$\mathbf{17.88(9.10)}$}& \textcolor{blue}{$\mathbf{26.34(9.12)}$}& \textcolor{blue}{$\mathbf{8.93(4.79)}$}& \textcolor{blue}{$\mathbf{207.60(195.20)}$}& \textcolor{blue}{$\mathbf{8.43(4.73)}$}& \textcolor{blue}{$\mathbf{1.36(0.97)}$} \\
\hline\hline
\end{tabular}
\label{tab:interp}
\end{table*}

\begin{table*}[t]
\caption{Mean(std) results of the reconstruction performance of different numbers of patches within the PMC component. The best results are shown in blue.}
\centering
\begin{tabular}{p{25pt}|p{40pt}|p{50pt}|p{50pt}|p{50pt}|p{70pt}|p{50pt}|p{45pt}}
\hline\hline
\multicolumn{1}{c|}{Dataset}& \multicolumn{1}{|c|}{Patch Number}& \multicolumn{1}{|c|}{FDR(\%)$\downarrow$}& \multicolumn{1}{|c|}{ADR(\%)$\downarrow$}& \multicolumn{1}{|c|}{MD(mm)$\downarrow$}& \multicolumn{1}{|c|}{SD(mm)$\downarrow$}& \multicolumn{1}{|c|}{HD(mm)$\downarrow$}& \multicolumn{1}{|c}{MEA(deg)$\downarrow$} \\
\hline
\multirow{4}*{Arm}& $8\times 8$& $11.68(6.43)$& $18.79(9.35)$& $29.38(15.75)$& $932.13(681.76)$& $26.61(12.53)$& $3.32(2.26)$ \\
~& $16\times 16$& $11.36(6.25)$& $18.41(9.21)$& $29.53(15.78)$& $905.84(661.16)$& $26.34(12.74)$& $3.14(2.26)$ \\
~& $32\times 32$& \textcolor{blue}{$\mathbf{11.04(6.02)}$}& \textcolor{blue}{$\mathbf{18.26(9.10)}$}& \textcolor{blue}{$\mathbf{28.84(15.62)}$}& \textcolor{blue}{$\mathbf{901.87(658.43)}$}& \textcolor{blue}{$\mathbf{26.24(12.47)}$}& \textcolor{blue}{$\mathbf{3.00(2.25)}$} \\
~& $64\times 64$& $11.12(6.01)$& $18.31(9.16)$& $28.99(15.65)$& $908.18(642.13)$& $26.25(12.58)$& $3.03(2.25)$ \\
\hline
\multirow{4}*{Carotid}& $8\times 8$& $12.54(7.92)$& $24.38(17.92)$& $21.53(9.21)$& $665.18(472.35)$& $19.05(8.91)$& $2.56(1.60)$ \\
~& $16\times 16$& $12.32(7.63)$& $24.07(17.65)$& $21.16(8.92)$& $650.85(465.91)$& $18.80(8.79)$& $2.49(1.58)$ \\
~& $32\times 32$& \textcolor{blue}{$\mathbf{12.11(7.55)}$}& \textcolor{blue}{$\mathbf{23.78(17.51)}$}& \textcolor{blue}{$\mathbf{20.75(8.82)}$}& \textcolor{blue}{$\mathbf{646.00(455.03)}$}& \textcolor{blue}{$\mathbf{18.50(8.64)}$}& \textcolor{blue}{$\mathbf{2.40(1.55)}$} \\
~& $64\times 64$& $12.20(7.64)$& $23.86(17.48)$& $20.89(8.90)$& $650.27(458.32)$& $18.64(8.66)$& $2.42(1.56)$ \\
\hline
\multirow{4}*{Thyroid}& $8\times 8$& $18.09(9.35)$& $26.85(9.53)$& $9.42(4.95)$& $216.43(202.18)$& $8.87(4.76)$& $1.43(1.00)$ \\
~& $16\times 16$& $17.98(9.15)$& $26.52(9.22)$& $9.05(4.81)$& $214.12(196.82)$& $8.51(4.74)$& $1.41(0.98)$ \\
~& $32\times 32$& \textcolor{blue}{$\mathbf{17.88(9.10)}$}& \textcolor{blue}{$\mathbf{26.34(9.12)}$}& \textcolor{blue}{$\mathbf{8.93(4.79)}$}& \textcolor{blue}{$\mathbf{207.60(195.20)}$}& \textcolor{blue}{$\mathbf{8.43(4.73)}$}& \textcolor{blue}{$\mathbf{1.36(0.97)}$} \\
~& $64\times 64$& $17.93(9.12)$& $26.45(9.18)$& $8.95(4.83)$& $209.33(198.76)$& $8.45(4.75)$& $1.39(0.97)$ \\
\hline\hline
\end{tabular}
\label{tab:patchsize}
\end{table*}

\appendix

\subsection{Comparison of Different Architectures in TMS}
\label{sec:results:resnet}

The TMS architecture consists of ResNet18 and ConvLSTM.
ResNet18 is highly effective in extracting spatial features, while ConvLSTM excels at capturing spatio-temporal information. 
While more complex temporal architectures such as Transformers~\cite{han2022survey} offer strong spatio-temporal learning capabilities, they have significantly larger numbers of parameters and require larger training datasets to perform optimally. This makes them less suitable for our task, where the dataset sizes are limited and computational resources are constrained.

To validate the effectiveness of the ResNet18 architecture in TMS, we conducted a comparison with two well-known architectures: ResNeXt50~\cite{xie2017aggregated} and MobileNetV4~\cite{qin2024mobilenetv4}. ResNeXt50 is a ResNet-based architecture designed with grouped convolutions to increase model capacity and improve feature representation efficiency. MobileNetV4 is an efficient architecture designed with universal inverted bottleneck blocks and neural architecture search for mobile and resource-constrained environments. 
In our comparison, we replaced ResNet18 with ResNeXt50 or MobileNetV4 in TMS.

Table~\ref{tab:resnet18} summarizes the overall comparison of ResNet18 with ResNeXt50 and MobileNetV4 on arm, carotid, and thyroid datasets. It can be seen that ResNet18 outperformed both ResNeXt50 and MobileNetV4 across all datasets and metrics. 
The lightweight design of ResNet18, combined with its strong feature extraction capabilities, contributed to this superior performance. In contrast, ResNeXt50's larger model capacity led to overfitting on our datasets, and while MobileNetV4 is efficient, it sacrifices some model capacity to maintain a smaller model size, which is not optimal for our task.
This result shows that ResNet18 achieves a better balance between feature extraction capability and computational efficiency to meet the specific requirements of our freehand 3D US reconstruction task.

\begin{table*}[t]
\caption{Mean(std) results of the reconstruction performance of different online loss weights. The best results are shown in blue.}
\centering
\begin{tabular}{p{25pt}|p{50pt}|p{50pt}|p{50pt}|p{50pt}|p{70pt}|p{50pt}|p{45pt}}
\hline\hline
\multicolumn{1}{c|}{Dataset}& \multicolumn{1}{|c|}{Weight of}& \multicolumn{1}{|c|}{FDR(\%)$\downarrow$}& \multicolumn{1}{|c|}{ADR(\%)$\downarrow$}& \multicolumn{1}{|c|}{MD(mm)$\downarrow$}& \multicolumn{1}{|c|}{SD(mm)$\downarrow$}& \multicolumn{1}{|c|}{HD(mm)$\downarrow$}& \multicolumn{1}{|c}{MEA(deg)$\downarrow$} \\
\multicolumn{1}{c|}{}& \multicolumn{1}{|c|}{SVC,PAC,PMC,MSS}& \multicolumn{1}{|c|}{}& \multicolumn{1}{|c|}{}& \multicolumn{1}{|c|}{}& \multicolumn{1}{|c|}{}& \multicolumn{1}{|c|}{}& \multicolumn{1}{|c}{} \\
\hline
\multirow{13}*{Arm}& 1, 1, 1, 1& \textcolor{blue}{$\mathbf{11.04(6.02)}$}& $18.26(9.10)$& $28.84(15.62)$& \textcolor{blue}{$\mathbf{901.87(658.43)}$}& \textcolor{blue}{$\mathbf{26.24(12.47)}$}& \textcolor{blue}{$\mathbf{3.00(2.25)}$} \\
\cline{2-8}
~&0.5, 1, 1, 1& $11.26(6.13)$& $18.32(9.31)$& $28.99(15.60)$& $908.63(654.32)$& $26.26(12.48)$& $3.03(2.26)$ \\
~&2, 1, 1, 1& $11.28(6.24)$& \textcolor{blue}{$\mathbf{18.24(9.12)}$}& $28.96(15.65)$& $909.45(660.19)$& $26.30(12.48)$& $3.02(2.26)$ \\
~&4, 1, 1, 1& $11.38(6.35)$& $18.39(9.29)$& $29.15(15.86)$& $916.42(659.27)$& $26.42(12.52)$& $3.05(2.27)$ \\
~&1, 0.5, 1, 1& $11.17(6.01)$& $18.38(9.24)$& $28.90(15.63)$& $905.32(658.09)$& $26.25(12.45)$& $3.02(2.25)$ \\
~&1, 2, 1, 1& $11.22(6.05)$& $18.32(9.16)$& $28.92(15.62)$& $906.19(663.17)$& $26.29(12.49)$& $3.01(2.25)$ \\
~&1, 4, 1, 1& $11.45(6.18)$& $18.47(9.29)$& $29.05(15.72)$& $912.18(678.70)$& $26.40(12.57)$& $3.03(2.26)$ \\
~&1, 1, 0.5, 1& $11.15(6.15)$& $18.30(9.19)$& $28.88(15.60)$& $908.89(654.37)$& $26.26(12.45)$& $3.01(2.25)$ \\
~&1, 1, 2, 1& $11.23(6.16)$& $18.37(9.24)$& $28.91(15.64)$& $903.81(660.21)$& $26.31(12.39)$& $3.01(2.25)$ \\
~&1, 1, 4, 1& $11.42(6.42)$& $18.51(9.46)$& $28.99(15.70)$& $910.75(665.19)$& $26.45(12.50)$& $3.03(2.26)$ \\
~&1, 1, 1, 0.5& $11.21(6.21)$& $18.36(9.05)$& \textcolor{blue}{$\mathbf{28.82(15.63)}$}& $904.81(657.09)$& $26.32(12.51)$& $3.02(2.25)$ \\
~&1, 1, 1, 2& $11.35(6.19)$& $18.31(9.10)$& $28.88(15.65)$& $909.15(651.16)$& $26.35(12.50)$& $3.02(2.26)$ \\
~&1, 1, 1, 4& $11.50(6.38)$& $18.46(9.27)$& $28.96(15.69)$& $918.52(665.27)$& $26.50(12.56)$& $3.06(2.27)$ \\
\hline
\multirow{13}*{Carotid}& 1, 1, 1, 1& \textcolor{blue}{$\mathbf{12.11(7.55)}$}& \textcolor{blue}{$\mathbf{23.78(17.51)}$}& $20.75(8.82)$& \textcolor{blue}{$\mathbf{646.00(455.03)}$}& \textcolor{blue}{$\mathbf{18.50(8.64)}$}& \textcolor{blue}{$\mathbf{2.40(1.55)}$} \\
\cline{2-8}
~&0.5, 1, 1, 1& $12.18(7.58)$& $23.85(17.53)$& $20.85(8.82)$& $651.16(466.16)$& $18.60(8.65)$& \textcolor{blue}{$\mathbf{2.40(1.55)}$} \\
~&2, 1, 1, 1& $12.15(7.56)$& $23.88(17.55)$& $20.86(8.93)$& $655.75(470.19)$& $18.56(8.62)$& $2.41(1.55)$ \\
~&4, 1, 1, 1& $12.20(7.60)$& $23.95(17.62)$& $20.93(8.99)$& $662.91(478.62)$& $18.75(8.70)$& $2.45(1.57)$ \\
~&1, 0.5, 1, 1& $12.14(7.58)$& $23.80(17.52)$& $20.80(8.85)$& $649.18(460.85)$& $18.56(8.64)$& $2.41(1.56)$ \\
~&1, 2, 1, 1& $12.19(7.62)$& $23.79(17.53)$& $20.82(8.84)$& $649.82(463.53)$& $18.59(8.66)$& $2.43(1.56)$ \\
~&1, 4, 1, 1& $12.23(7.65)$& $23.86(17.60)$& $20.88(8.92)$& $655.75(472.19)$& $18.70(8.72)$& $2.46(1.58)$ \\
~&1, 1, 0.5, 1& $12.15(7.58)$& $23.81(17.60)$& $20.81(8.87)$& $648.02(459.82)$& $18.53(8.63)$& $2.42(1.56)$ \\
~&1, 1, 2, 1& $12.15(7.54)$& $23.80(17.58)$& $20.79(8.76)$& $649.35(461.76)$& $18.53(8.65)$& \textcolor{blue}{$\mathbf{2.40(1.55)}$} \\
~&1, 1, 4, 1& $12.19(7.60)$& $23.88(17.60)$& $20.87(8.83)$& $652.66(470.60)$& $18.68(8.70)$& $2.45(1.57)$ \\
~&1, 1, 1, 0.5& $12.17(7.56)$& $23.82(17.58)$& \textcolor{blue}{$\mathbf{20.72(8.80)}$}& $649.92(460.37)$& $18.56(8.68)$& $2.43(1.56)$ \\
~&1, 1, 1, 2& $12.19(7.58)$& $23.81(17.60)$& $20.76(8.82)$& $653.42(465.81)$& $18.60(8.72)$& $2.44(1.57)$ \\
~&1, 1, 1, 4& $12.26(7.63)$& $23.89(17.65)$& $20.88(8.86)$& $659.05(474.06)$& $18.76(8.77)$& $2.48(1.58)$ \\
\hline
\multirow{13}*{Thyroid}& 1, 1, 1, 1& \textcolor{blue}{$\mathbf{17.88(9.10)}$}& \textcolor{blue}{$\mathbf{26.34(9.12)}$}& \textcolor{blue}{$\mathbf{8.93(4.79)}$}& $207.60(195.20)$& \textcolor{blue}{$\mathbf{8.43(4.73)}$}& \textcolor{blue}{$\mathbf{1.36(0.97)}$} \\
\cline{2-8}
~&0.5, 1, 1, 1& $17.93(9.15)$& $26.38(9.15)$& $8.96(4.82)$& $212.54(198.16)$& $8.45(4.72)$& $1.38(0.97)$ \\
~&2, 1, 1, 1& $17.96(9.18)$& $26.42(9.17)$& $9.03(4.89)$& $215.10(199.36)$& $8.45(4.73)$& $1.38(0.97)$ \\
~&4, 1, 1, 1& $18.06(9.32)$& $26.50(9.25)$& $9.12(4.95)$& $220.41(205.42)$& $8.50(4.75)$& $1.40(0.98)$ \\
~&1, 0.5, 1, 1& $17.92(9.14)$& $26.38(9.19)$& $8.95(4.84)$& $208.15(202.75)$& $8.46(4.74)$& $1.37(0.96)$ \\
~&1, 2, 1, 1& $17.89(9.11)$& $26.40(9.23)$& $9.02(4.90)$& \textcolor{blue}{$\mathbf{206.92(193.16)}$}& $8.44(4.72)$& \textcolor{blue}{$\mathbf{1.36(0.97)}$} \\
~&1, 4, 1, 1& $18.03(9.29)$& $26.48(9.28)$& $9.08(4.92)$& $217.43(204.83)$& $8.51(4.80)$& $1.40(0.98)$ \\
~&1, 1, 0.5, 1& $17.90(9.15)$& $26.41(9.16)$& $8.96(4.82)$& $209.16(201.06)$& $8.46(4.75)$& $1.37(0.97)$ \\
~&1, 1, 2, 1& $17.95(9.17)$& $26.45(9.23)$& $8.98(4.80)$& $208.54(201.85)$& $8.45(4.74)$& $1.37(0.96)$ \\
~&1, 1, 4, 1& $18.09(9.35)$& $26.53(9.38)$& $9.06(4.90)$& $213.64(204.14)$& $8.49(4.76)$& $1.39(0.98)$ \\
~&1, 1, 1, 0.5& $17.91(9.16)$& $26.45(9.16)$& $8.99(4.81)$& $209.14(203.60)$& $8.44(4.72)$& $1.38(0.97)$ \\
~&1, 1, 1, 2& $17.97(9.19)$& $26.48(9.26)$& $9.01(4.89)$& $211.06(203.75)$& $8.45(4.73)$& $1.37(0.97)$ \\
~&1, 1, 1, 4& $18.04(9.31)$& $26.60(9.40)$& $9.13(4.96)$& $221.43(206.31)$& $8.52(4.74)$& $1.42(0.99)$ \\
\hline\hline
\end{tabular}
\label{tab:loss}
\end{table*}

\subsection{Comparison of Different Integration Methods of Velocity Feature in TMS}
\label{sec:results:integrate}

We analyze the different integration methods of the velocity feature in TMS, including addition and concatenation. As shown in Table~\ref{tab:integrate}, the addition method outperforms concatenation in terms of reconstruction performance, reducing FDR/ADR/MEA metrics by $6.71\%/4.60\%/0.23\%$, $3.54\%/0.98\%/2.72\%$, and $8.03\%/8.29\%/2.19\%$ across the arm, carotid, and thyroid datasets, respectively. This result confirms that the effectiveness of adding the velocity feature with the image feature.

\subsection{Ablation Study on Interpolated Images in PMC}
\label{sec:results:interpolated}
We conducted an ablation study on the interpolated images within the PMC component. As shown in Table~\ref{tab:interp}, the inclusion of interpolated images (number$>$0) has better reconstruction performance than the exclusion of interpolated images (number$=$0) across all datasets. This result confirms the effectiveness of PMC's interpolated images in capturing inter-frame nonlinear changes from complex motion trajectories. Furthermore, the reconstruction performance improves as the number of interpolated images increases across all datasets. Considering the computational cost of interpolation, we chose to set the number of interpolated images to 63.

\subsection{Comparison of Different Patch Numbers in PMC}
\label{sec:results:patchsize}

We analyze the different numbers of patches ($P\times Q$) within the PMC component, where $P$ and $Q$ determine the numbers of rows and columns of patches in the image, respectively. As shown in Table~\ref{tab:patchsize}, the $32\times 32$ patch number outperforms other numbers in terms of reconstruction performance across all datasets. This result shows that the $32\times 32$ patch number strikes an optimal balance between capturing fine-grained motion variations and avoiding pixel-level speckle noise. Smaller patch sizes (larger number of patches) tend to focus too narrowly on image details, potentially emphasizing speckle noise rather than actual motion, while larger patch sizes (smaller number of patches) may fail to capture fine-grained motion variations.

\subsection{Comparison of Different Online Loss Weights}
\label{sec:results:lossweight}

We analyze the different weights of the four online losses ($L_{SVC}$, $L_{PAC}$, $L_{PMC}$, and $L_{MSS}$) using the control variable method. Specifically, we assign different weights ($0.5$, $2$, and $4$) to each loss while setting the weights of other losses to 1. As shown in Table~\ref{tab:loss}, MoNetV2 achieves the best performance across all datasets and nearly all metrics when all weights are set to 1. Compared to other weights, deviations from equal weighting result in slight performance drops across various metrics. This confirms that assigning equal weights to all losses ensures balanced optimization, preventing any single loss from dominating the online process.

\subsection{Analysis of Scanning Direction Changes, In-Plane and Out-of-Plane Displacements, and Dihedral Angles}
\label{sec:results:analysis}

We analyze the performance of several aspects within trajectory estimation, including scanning direction changes, in-plane and out-of-plane displacements, and dihedral angles, to reflect the differences in trajectory estimation between different methods. 
For the detection of scanning direction changes, we utilize the angle between translations in adjacent inter-frame transformations to determine whether the scanning direction has changed on a particular frame in the scan. 
Due to the sparsity of direction-changing frames and the similarity of content between adjacent frames, metrics should not rigidly demand an exact match between the estimated direction-changing frame and the actual one. Generally, the closer the estimated index is to the actual index, the more accurate the detection. Inspired by the Top-K metric, we define Precision-K, Recall-K, and F1-score-K metrics based on precision, recall, and the F1-score metrics to evaluate the accuracy of direction-changing frame detection. Precision-K, Recall-K, and F1-score-K exhibit loosened constraints by allowing the distance between the estimated and actual indices to be smaller than a given threshold $K$. For a real index set $S$ and the estimated index set $\hat{S}$ of direction-changing frames, the metrics can be expressed as:
\begin{equation}
\left\{
\begin{aligned}
&\textit{Precision-K}=\frac{\textit{TP-K}}{\left|\hat{S}\right|}, \\
&\textit{Recall-K}=\frac{\textit{TP-K}}{\left|S\right|}, \\
&\textit{F1-score-K}=\frac{2\times\textit{Precision-K}\times\textit{Recall-K}}{\textit{Precision-K}+\textit{Recall-K}},
\end{aligned}
\right.
\end{equation}
where $|S|$ and $|\hat{S}|$ represent the cardinal numbers of the $S$ and $\hat{S}$, respectively.
\begin{equation}
\textit{TP-K}=\sum_{i\in S_{min}}\textit{sgn}\left(K-\left|R(i)-i\right|\right).
\end{equation}
$R(\cdot)$ is an injective function $R:S_{min}\xrightarrow{}S_{max}$ and minimizes the value of $\sum_{i\in S_{min}}|R(i)-i|$, where $S_{min}$ and $S_{max}$ represent the sets with fewer and more elements between $S$ and $\hat{S}$, respectively. $\textit{sgn}(\cdot)$ denotes the sign function:
\begin{equation}
\textit{sgn}(x)=
\left\{
\begin{aligned}
&1,\quad x\ge 0, \\
&0,\quad otherwise.
\end{aligned}
\right.
\end{equation}

\begin{figure}[t]
\centerline{\includegraphics[width=\columnwidth]{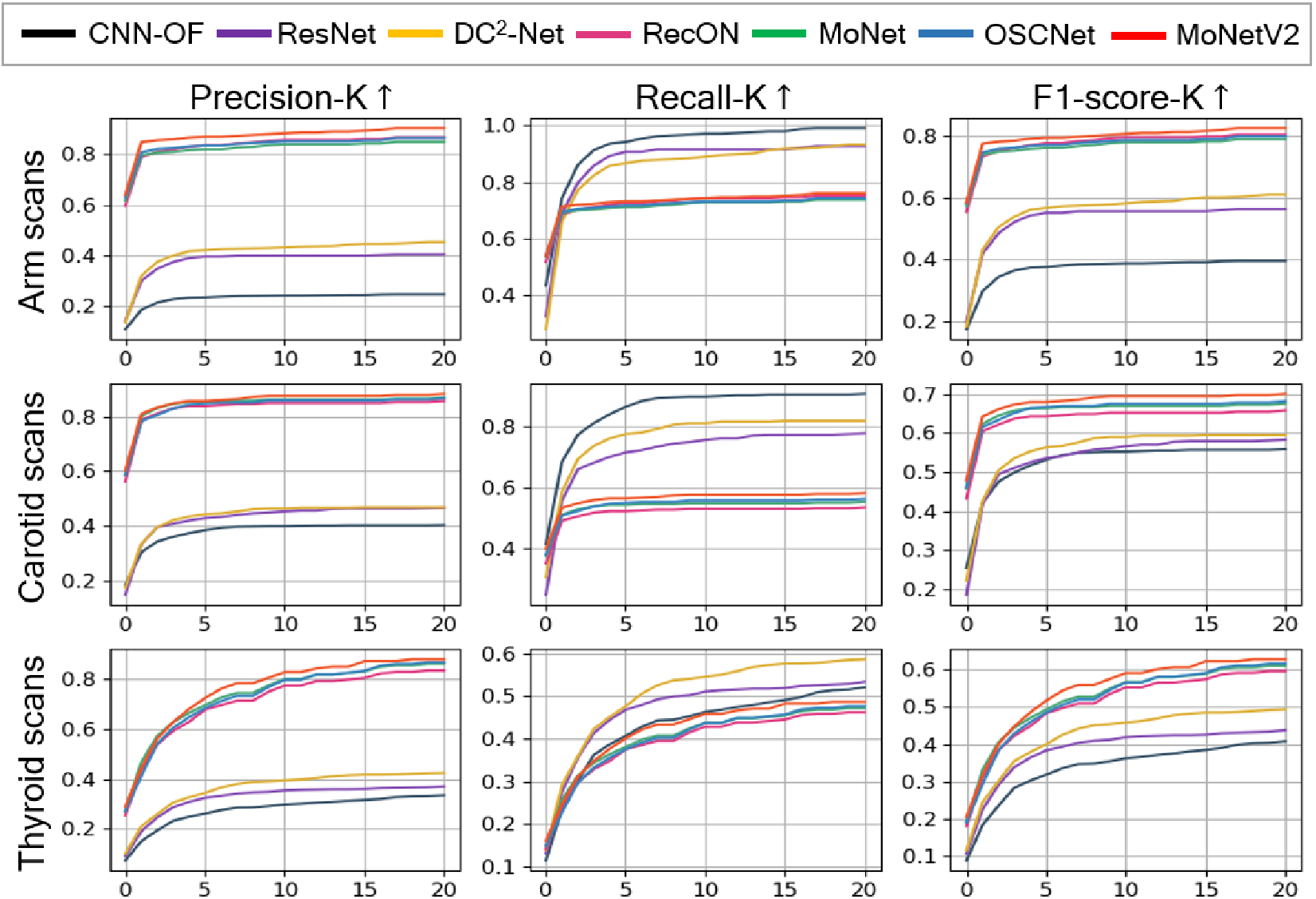}}
\caption{Precision-K, Recall-K, and F1-score-K curves on arm (Row I), carotid (Row II), and thyroid (Row III) scans. The x- and y-axis show the $K$ value and metrics, respectively.}
\label{fig11}
\end{figure}

Fig.~\ref{fig11} shows the Precision-K, Recall-K, and F1-score-K curves of all methods. As $K$ increases, all metrics increase significantly and tend to converge. It can be observed that RecON, MoNet, OSCNet, and MoNetV2 have relatively high Precision-K values (e.g., MoNetV2's value converges to 0.9), indicating high accuracy in predicting frames undergoing changes in direction. An interesting phenomenon is that CNN-OF, ResNet, and DC$^2$-Net have relatively high Recall-K values. This is due to them incorrectly predicting many frames that have not experienced a change in direction as direction-changing frames. For instance, CNN-OF predicts $30\%$ of the frames in all scans as direction-changing frames. Precision-K and Recall-K as individual metrics are limited. F1-score-K is the harmonic mean of Precision-K and Recall-K, providing a balance between these two metrics. As shown in Fig.~\ref{fig11}, MoNetV2 obtains the highest F1-score-K curves in all three datasets, which indicates that MoNetV2 has an advantage over other methods in direction-changing frame detection. 

\begin{figure}[t]
\centerline{\includegraphics[width=\columnwidth]{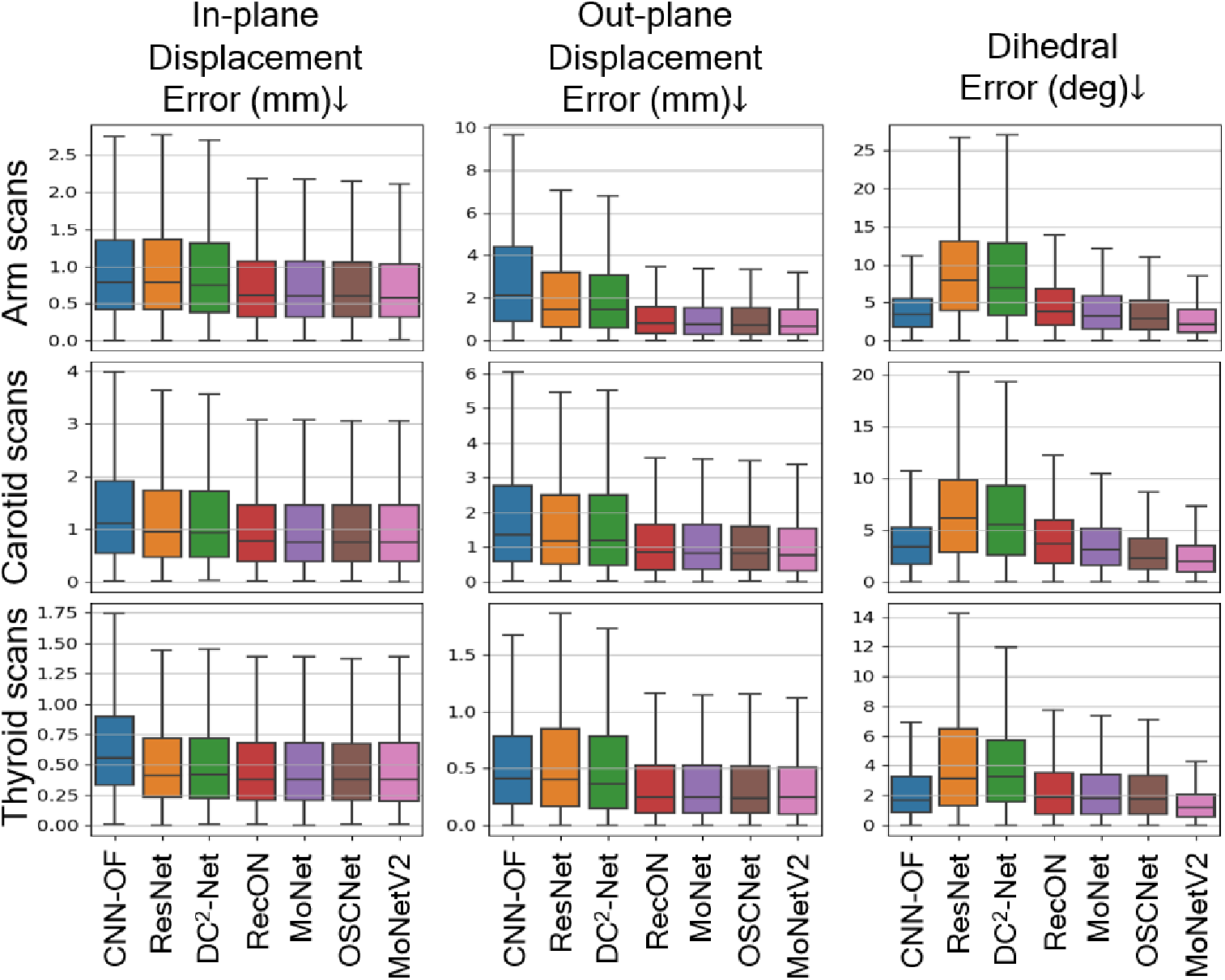}}
\caption{Error boxplots for the in-plane displacement, out-of-plane displacement, and dihedral angle on arm (Row I), carotid (Row II), and thyroid (Row III) scans. Each boxplot includes minimum, first quartile (Q1), median, third quartile (Q3), and maximum values. The x- and y-axis show the methods and errors, respectively.}
\label{fig12}
\end{figure}

We analyze the in-plane displacement, out-of-plane displacement, and dihedral angle, as shown in Fig.~\ref{fig12}. The errors of the in-plane and out-of-plane displacements can be calculated as $\sqrt{(\hat{t}_x-t_x)^2+(\hat{t}_y-t_y)^2}$ and $|\hat{t}_z-t_z|$, respectively. In which $\hat{t}=(\hat{t}_x,\hat{t}_y,\hat{t}_z)$ and $t=(t_x,t_y,t_z)$ denote the estimated and actual translation, respectively. The error of the dihedral angle can be expressed as $\cos^{-1}{(\hat{\vec{n}}\cdot \vec{n})}$, where $\hat{\vec{n}}$ and $\vec{n}$ denote the estimated and actual normal vectors of the frames in the scan, respectively. Fig.~\ref{fig12} shows that MoNetV2 outperforms all other methods across all three datasets. This indicates that MoNetV2 has minimal in-plane or out-of-plane drift originating from accumulated errors. 
In addition, the advantages of RecON, MoNet, OSCNet, and MoNetV2 in terms of displacement and angle errors, as well as the advantage of CNN-OF in terms of angle errors, again illustrate the necessity of motion information integration.

\subsection{Analysis of Inference Time}
\begin{figure}[h]
\centerline{\includegraphics[width=\columnwidth]{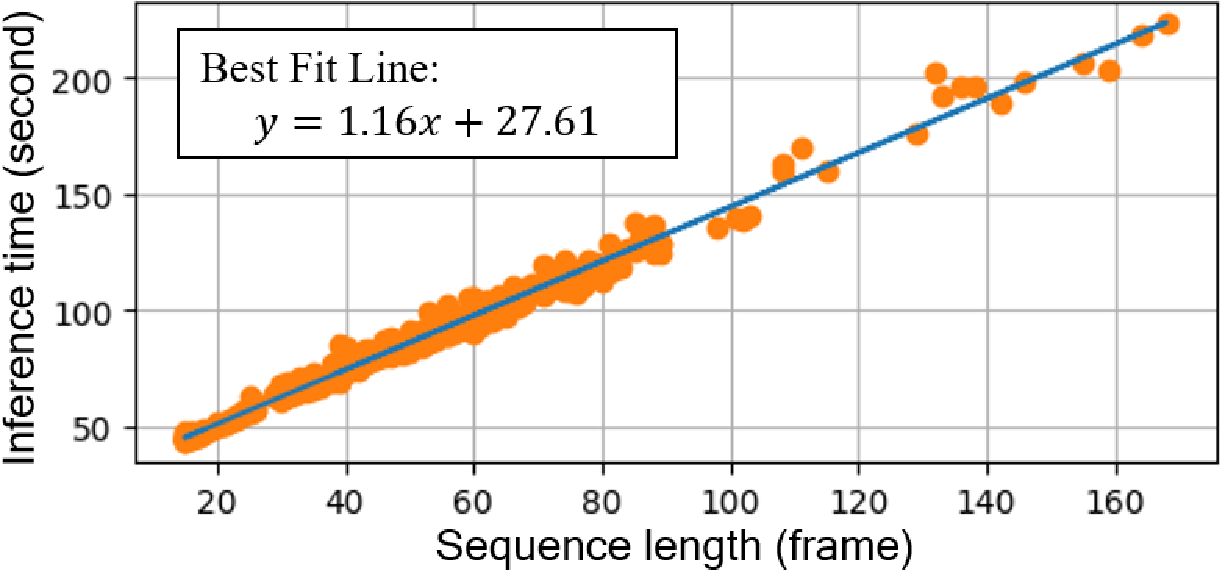}}
\caption{Correlation of the sequence lengths and inference times. Each orange dot indicates a sample. The blue line represents the line of best fit for all samples.}
\label{fig:time}
\end{figure}

We analyze the average inference time for all methods. All methods were run on a server equipped with RTX 3090 GPUs, and the results are shown in Table~\ref{tab:time}. It can be seen that the online-learning-based methods (RecON, MoNet, OSCNet, and MoNetV2) exhibit longer average inference times compared to the others (CNN-OF, ResNet, and DC$^2$-Net). This is due to the adaptive optimization in the test phase by online learning. However, MoNetV2 achieves faster inference than RecON and OSCNet by excluding computationally intensive 3D operations and utilizing a single IMU sensor. 
Compared to MoNet, MoNetV2 is slower due to the introduction of several new loss terms for motion consistency (SVC, PAC, and PMC). While these increase the computational requirements during both the forward pass and gradient updates, they also significantly improve reconstruction accuracy.
As a result, MoNetV2 offers an efficient trade-off, delivering relatively faster inference with better accuracy and adaptability. Fig.~\ref{fig:time} illustrates the strong correlation between sequence length and inference time for MoNetV2.

\begin{table}[h]
\caption{Average inference time for different methods.}
\centering
\begin{tabular}{c|c}
\hline\hline
Method & Average Inference Time (second) \\
\hline
CNN-OF & 0.1456 \\
ResNet & 0.1486 \\
DC$^2$-Net & 1.5107 \\
RecON & 611.2099 \\
MoNet & 44.7535 \\
OSCNet & 217.1149 \\
MoNetV2 & 87.8272 \\
\hline\hline
\end{tabular}
\label{tab:time}
\end{table}

\bibliographystyle{elsarticle-num}
\bibliography{refs}

\end{document}